\documentclass[prd,showpacs,preprint]{revtex4}
\usepackage{amssymb}
\usepackage{amsmath}
\usepackage{graphicx}
\usepackage{multirow}
\usepackage{subfigure}
\usepackage{float}

\begin{document}

\title{Next-to-leading order QCD corrections to the single
top quark production via model-independent t-q-g flavor-changing
neutral-current couplings at
hadron colliders}
\author{Jun Gao}
\author{Chong Sheng Li}
\email{csli@pku.edu.cn}
\author{Jia Jun Zhang}
\author{Hua Xing Zhu}
\affiliation{Department of Physics and State Key Laboratory of
Nuclear Physics and Technology, Peking University, Beijing 100871,
China}

\date{\today}


\begin{abstract}
We present the calculations of the complete
next-to-leading order (NLO) QCD effects on the single top
productions induced by model-independent $tqg$ flavor-changing
neutral-current couplings at hadron colliders. Our results show
that, for the $tcg$ coupling the NLO QCD corrections can enhance the
total cross sections by about 60\% and 30\%, and for the $tug$
coupling by about 50\% and 20\% at the Tevatron and LHC,
respectively, which means that the NLO corrections can increase the
experimental sensitivity to the FCNC couplings by about
10\%$-$30\%. Moreover, the NLO corrections reduce the dependence
of the total cross sections on the renormalization or factorization
scale significantly, which lead to increased confidence on the
theoretical predictions. Besides, we also evaluate the NLO
corrections to several important kinematic distributions, and find
that for most of them the NLO corrections are almost the same and do
not change the shape of the distributions.
\end{abstract}

\maketitle

\section{Introduction}\label{s1}
The top quark is the heaviest particle so far discovered, with a
mass close to the electroweak (EW) symmetry breaking scale. Thus it
is a wonderful probe for the EW breaking mechanism and new physics
beyond the standard model (SM) through its decays and productions at
colliders. One of the most important aspects of the top quark physics
is the investigation of the anomalous flavor-changing neutral-currents
(FCNC) couplings in the top quark sector. Within the SM, the FCNC
couplings are absent at the tree level, and occur at high order in
perturbation theory through loop diagrams which are further
suppressed by the Glashow-Iliopoulos-Maiani (GIM)
mechanism~\cite{Glashow:1970gm}. On the other hand, these FCNC
couplings can be enhanced to observable levels in some new physics
models~\cite{AguilarSaavedra:2004wm}, such as the models with extra
quarks~\cite{delAguila:1998tp}, two Higgs doublet
models~\cite{Cheng:1987rs}, supersymmetric models~\cite{Li:1993mg},
extra dimensions models~\cite{Davoudiasl:2001uj}, little Higgs
models~\cite{HongSheng:2007ve}, or technicolor
models~\cite{Wang:1994qd}. As the coming Large Hadron Collider (LHC)
will produce abundant top quark events (about $10^8$ per year), even
in the initial low luminosity run ($\sim 10{\rm \ fb}^{-1}$/year)
$8\times10^6$ top quark pairs and $3\times10^6$ single top quarks
will be produced yearly, one may anticipate the discovery of the first
hint of new physics by observing the FCNC couplings in the top quark
sector.

Since we do not know which type of new physics will be responsible
for a future deviation from the SM predictions, it is necessary to
study the top quark FCNC processes in a model-independent way by an
effective Lagrangian. In general, any new physics at a high scale
$\Lambda$ will manifest themselves at energies below $\Lambda$
through small deviations from the SM, which can be described by an
effective Lagrangian containing higher dimensional SM gauge
invariant operators~\cite{Buchmuller:1985jz}. As we know, the
dimension five operators break baryon and lepton number
conservation, the lowest order operators considered are dimension
six. For the $tqg$ anomalous FCNC couplings, the only independent
one is the chromo-magnetic operator induced by a dimension six
gauge invariant operator before spontaneous gauge symmetry
breaking~\cite{AguilarSaavedra:2008zc},
\begin{equation}\label{eq0}
g_s\sum_{q=u,c}\frac{\kappa^g_{tq}}{\Lambda}\bar{t}\sigma_{\mu\nu}
T^a q G^a_{\mu\nu}+H.c.,
\end{equation}
where $T^a$ are the Gell-Mann matrices, $G^a_{\mu\nu}$ are the field
strength tensors of the gluon, and $\kappa^g_{tq}\ (q=u,c)$ are real
coefficients that define the strength of the couplings. To test the
above couplings at the Tevatron and LHC, the direct top quark
production~\cite{Hosch:1997gz} and the FCNC single top quark
production~\cite{Malkawi:1995dm,Han:1998tp} are the most promising processes, and the
top quark rare decay process is not so efficient compared with the above two
processes due to the large SM backgrounds~\cite{Carvalho:2007yi}.
Currently the most stringent experimental constraints for the $tqg$
anomalous couplings are $\kappa^g_{tu}/\Lambda\leq 0.018\ {\rm
TeV^{-1}}$ and $\kappa^g_{tc}/\Lambda\leq 0.069\ {\rm TeV^{-1}}$,
given by the CDF collaboration through the measurements of the
direct top quark production~\cite{Aaltonen:2008qr} using
$2.2\ {\rm fb^{-1}}$ of data. The D0 collaboration also analyzes
$230\ {\rm pb^{-1}}$ of data and provides similar constraints,
$\kappa^g_{tu}/\Lambda\leq 0.037\ {\rm TeV^{-1}}$ and
$\kappa^g_{tc}/\Lambda\leq 0.15\ {\rm TeV^{-1}}$, based on the
measurements of the FCNC single top production~\cite{Abazov:2007ev}
using the next-to-leading order (NLO) K factors of the direct top
quark production process~\cite{Liu:2005dp}. Taking into account the
difference of the amount of data, we can see that the discovery
potential of these two channels are almost the same at the Tevatron.
As for the LHC, according to the tree-level analysis in
Ref.~\cite{Han:1998tp}, for the FCNC single top quark production,
$\kappa^g_{tq}/\Lambda$ can be detected
down to $0.0061\ {\rm TeV^{-1}}$ and $0.013\ {\rm TeV^{-1}}$ with
an integrated luminosity of $10\ {\rm fb^{-1}}$
for $q=u$ and $c$, respectively, thus it is one
of the most important channels to detect the top FCNC couplings
at the LHC.

As we know, the leading order (LO) cross sections for processes at
hadron colliders suffer from large uncertainties due to the
arbitrary choice of the renormalization scale ($\mu_r$) and
factorization scale ($\mu_f$), thus are not sufficient for the
extraction of the FCNC couplings. Then high order corrections
including NLO QCD and resummation effects are needed in order to
improve the theoretical predictions. The NLO QCD and threshold
resummation effects for the direct top quark production are studied
in Ref.~\cite{Liu:2005dp}, and the NLO QCD corrections to the top
quark rare decays via FCNC couplings can be found in
Ref.~\cite{Zhang:2008yn}. As for the FCNC single top production,
all the predictions are still at the LO. In this paper,
we present the complete NLO QCD corrections to the single top quark
production via model-independent $tqg$ FCNC couplings at
both the Tevatron and LHC.

The arrangement of this paper is as follows. In Sec.~\ref{s2} we
show the LO results. In Sec.~\ref{s3}, we present the details of the
NLO calculations, including the virtual and real corrections.
Section~\ref{s4} contains the numerical results, and
Section~\ref{s5} is a brief summary.

\section{Leading order results}\label{s2}
\begin{figure}[h]
\begin{center}
\scalebox{0.5}{\includegraphics*[40,420][750,740]{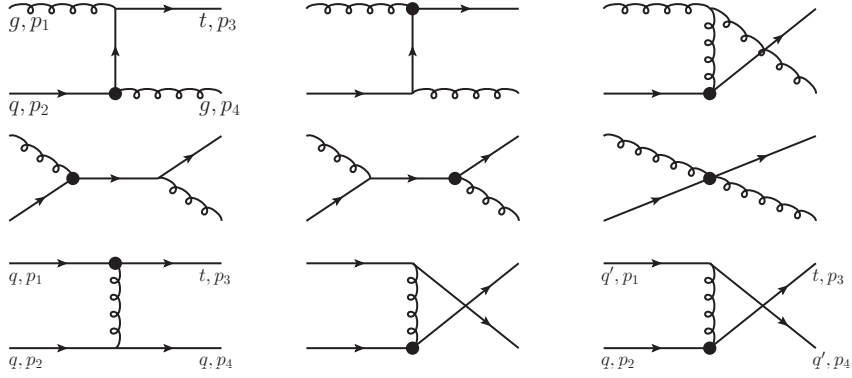}}
\caption[]{\label{f1}The LO Feynman diagrams for the single top
quark production via the FCNC couplings. The ones for $gg$ and $\bar
q q$ subprocesses can be obtained by crossing symmetry.}
\end{center}
\end{figure}

At the LO there are three main subprocesses which contribute to the
single top production via the FCNC couplings at hardron colliders:
\begin{equation}
g \ q \longrightarrow t \ g ,\ g \ g \longrightarrow t\ \bar{q}, \
q(\bar{q},q')\ q \longrightarrow t\ q(\bar{q},q'),
\end{equation}
where $q$ is either $u$ quark or $c$ quark. The corresponding
Feynman diagrams are shown in Fig.~\ref{f1}. It should be noted
that, after expanding the effective operator in Eq.~(\ref{eq0}),
there are two kinds of FCNC vertices, one is the three point vertex,
another is the four point vertex, and both of them contribute to the
subprocesses at the LO.

The LO squared amplitudes for the corresponding subprocesses in four
dimensions are
\begin{eqnarray}\label{eq1}
\overline{|M^B|^2}_{gq}(s,t,u)&=& \frac{\lambda^2}{9
   s t u \left(m^2-s\right)^2 \left(m^2-t\right)^2}
   \left(4 (s+t) m^{14}-\left(8 s^2+37 t s +8
   t^2\right)m^{12}\right.\nonumber \\
   &&+\left(8 s^3+92 t s^2+92 t^2 s+8
   t^3\right) m^{10}-2 \left(4 s^4+53 t s^3+109 t^2 s^2\right.\nonumber \\
   &&\left.+53 t^3 s+4 t^4\right) m^8
   +\left(4 s^5+66 t s^4+234 t^2s^3+234 t^3 s^2+66t^4 s\right.\nonumber \\
   &&\left. +4 t^5\right) m^6-s t \left(19 s^4+
   122 t s^3+221 t^2 s^2+122 t^3 s+19 t^4\right)m^4\nonumber \\
   &&\left.+6s^2 t^2 \left(5 s^3+14 t s^2
   +14 t^2 s+5 t^3\right) m^2-3s^3 t^3
   \left(5 s^2+4 t s+5 t^2\right)\right),\nonumber \\
\overline{|M^B|^2}_{qq}(s,t,u)&=& \frac{-4\lambda^2}{27t u} \left(3
m^6-3 (3 s+2 t) m^4 +2 \left(6 s^2+8 t s+3 t^2\right) m^2
- 2 s \left(3 s^2\right.\right.\nonumber \\
&&\left. \left.+5 t s+5t^2\right)\right),\nonumber \\
\overline{|M^B|^2}_{q'q}(s,t,u)&=& \frac{-4\lambda^2}{9 u}\left(m^2
(s+t)-2 s t\right), \nonumber \\
\overline{|M^B|^2}_{gg}(s,t,u)&=& -\frac{3}{8}\
\overline{|M^B|^2}_{gq}(u,t,s),\quad
\overline{|M^B|^2}_{\bar{q}q}(s,t,u)=\overline{|M^B|^2}_{qq}(t,s,u),
\end{eqnarray}
respectively, where $m$ is the top quark mass, and
$\lambda=8\pi\alpha_s \kappa^g_{tq}/\Lambda$, the colors and spins
of the outgoing particles have been summed over, and the colors and
spins of the incoming ones have been averaged over, $s$, $t$, and
$u$ are Mandelstam variables, which are defined as
\begin{equation}
s=(p_1+p_2)^2,\ \ t=(p_1-p_3)^2,\ \ u=(p_1-p_4)^2.
\end{equation}
After the phase space integration, the LO partonic cross sections
are given by
\begin{equation}
\hat \sigma^B_{ab}=\frac{1}{2\hat s}\int d\Gamma
\overline{|M^B|^2}_{ab}.
\end{equation}
The LO total cross section at hadron colliders is obtained by
convoluting the partonic cross section with the parton distribution
functions (PDFs) $G_{i/P}$ for the proton (antiproton):
\begin{equation}
\sigma^B=\sum_{ab}\int dx_1
dx_2\left[G_{a/P_1}(x_1,\mu_f)G_{b/P_2}(x_2,\mu_f)\hat
\sigma^B_{ab}\right],
\end{equation}
where $\mu_f$ is the factorization scale.

\section{Next-to-leading order QCD corrections}\label{s3}

The NLO corrections to the single top production via the FCNC
couplings consist of the virtual corrections, generated by loop
diagrams of colored particles, and the real corrections with the
radiation of a real gluon or a massless (anti)quark. We carried out
all the calculations in the 't Hooft-Feynman gauge and used the
dimensional regularization (DREG) scheme~\cite{'tHooft:1972fi} in
$n=4-2\epsilon$ dimensions to regularize all the divergences.
Moreover, for the real corrections, we used the dipole subtraction
method with massive partons~\cite{Catani:2002hc} to separate the
infrared (IR) divergences, which is convenient for the case of
massive Feynman diagrams and provides better numerical accuracy.

\subsection{Virtual corrections}

\begin{figure}[h]
\begin{center}
\scalebox{0.5}{\includegraphics*[70,8][840,800]{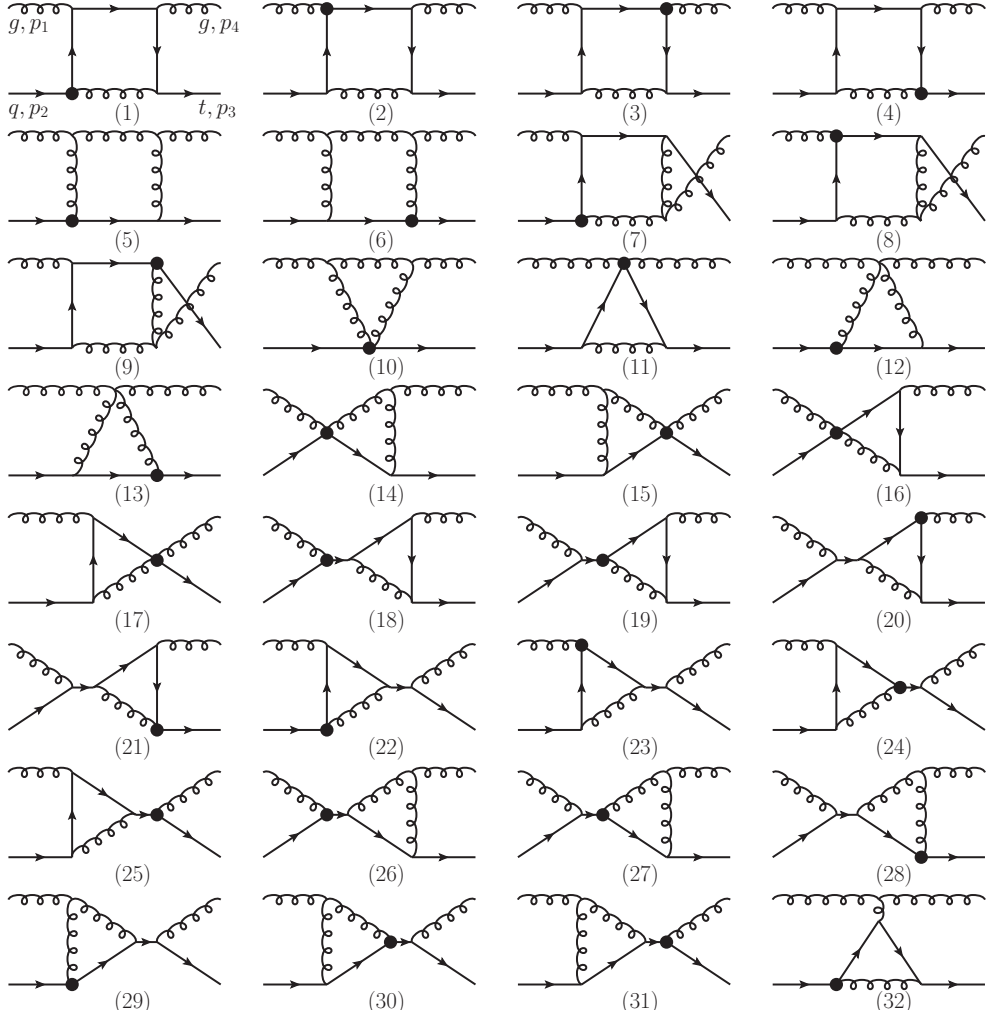}}
\caption[]{\label{f2}One-loop Feynman diagrams for the subprocess
$gq\to tg$, part I.}
\end{center}
\end{figure}

\begin{figure}[h]
\begin{center}
\scalebox{0.5}{\includegraphics*[70,10][840,800]{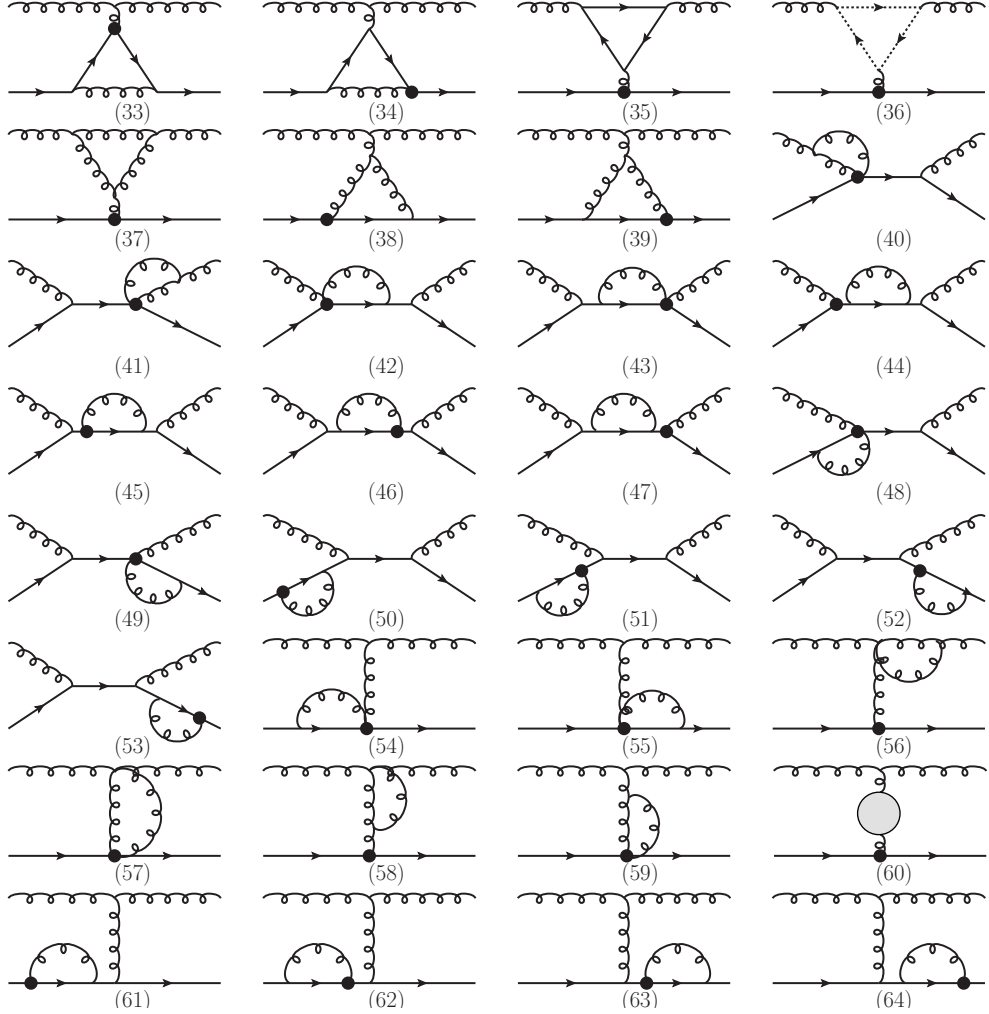}}
\caption[]{\label{f3}One-loop Feynman diagrams for the subprocess
$gq\to tg$, part II. The big gray circle represents the gluon
self-energy diagrams.}
\end{center}
\end{figure}

\begin{figure}[h]
\begin{center}
\scalebox{0.5}{\includegraphics*[70,245][840,750]{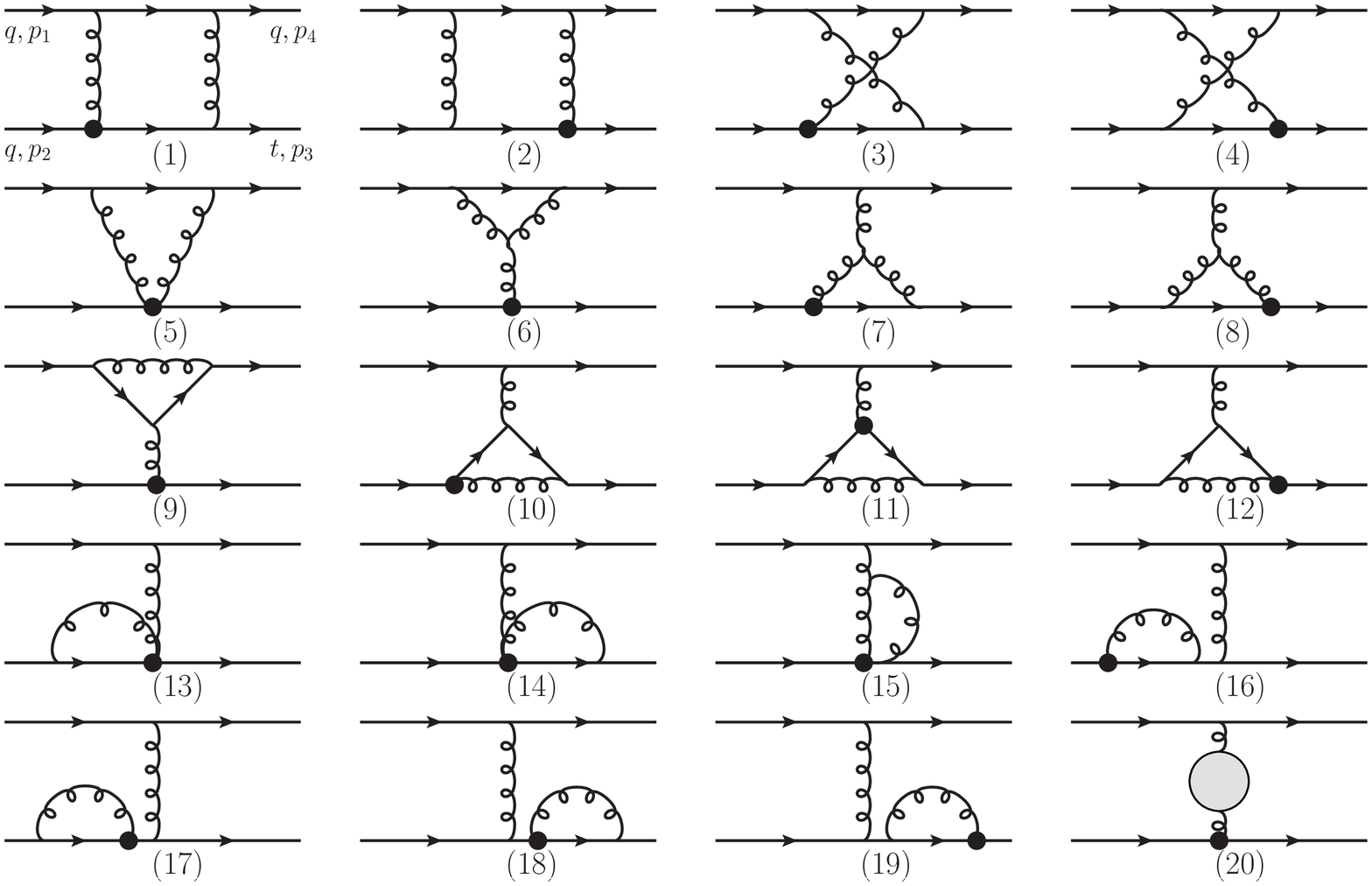}}
\caption[]{\label{f4}One-loop Feynman diagrams for the subprocess
$qq\to tq$.}
\end{center}
\end{figure}

The virtual corrections for the single top production via the FCNC
couplings include the box diagrams, triangle diagrams and
self-energy diagrams as shown in Figs.~\ref{f2}-\ref{f3} and
Fig.~\ref{f4} for the $gq$ and $qq$ initial state subprocess,
respectively. For simplicity we did not show the diagrams that only
differ by the exchange of identical external particles. The loop
diagrams for the $gg$ and $\bar q q$ initial state subprocess can be
obtained from Figs.~\ref{f2}-\ref{f3} and Fig.~\ref{f4} by crossing
symmetry, and the ones for $q'q$ are just part of Fig.~\ref{f4}.

All the ultraviolet (UV) divergences appearing in the loop diagrams
are renormalized by introducing counterterms for the wave functions
and mass of the external fields ($\delta Z_2^{(g)},\delta
Z_2^{(q)},\delta Z_2^{(t)},\delta m$), and the coupling constants
($\delta Z_{g_s},\delta Z_{\kappa^g_{tq}/\Lambda}$). We define these
counterterms according to the following conventions. For the
external fields, we fix all the renormalization constants using
on-shell subtraction, and, therefore, they also have IR
singularities:
\begin{eqnarray}
\delta Z_2^{(g)}&=&-\frac{\alpha_s}{2\pi}C_{\epsilon}\left(
\frac{n_f}{3}-\frac{5}{2}\right)\left(\frac{1}{\epsilon_{UV}}-
\frac{1}{\epsilon_{IR}}\right)-\frac{\alpha_s}{6\pi}C_{\epsilon}
\frac{1}{\epsilon_{UV}}, \nonumber \\
\delta Z_2^{(q)}&=&-\frac{\alpha_s}{3\pi}C_{\epsilon}\left(\frac{1}
{\epsilon_{UV}}-\frac{1}{\epsilon_{IR}}\right), \nonumber \\
\delta Z_2^{(t)}&=&-\frac{\alpha_s}{3\pi}C_{\epsilon}\left(\frac{1}
{\epsilon_{UV}}+\frac{2}{\epsilon_{IR}}+4\right), \nonumber \\
\frac{\delta m}{m}&=&-\frac{\alpha_s}{3\pi}C_{\epsilon}\left(\frac
{3}{\epsilon_{UV}}+4\right),
\end{eqnarray}
where $C_{\epsilon}=\Gamma(1+\epsilon)[(4\pi\mu_r^2)/m^2]
^{\epsilon}$ and $n_f=5$. For the renormalization of $g_s$, we use
the $\rm \overline{MS}$ scheme modified to decouple the top
quark~\cite{Collins:1978wz}, i.e. the first $n_f$ light flavors are
subtracted using the $\rm \overline{MS}$ scheme, while the
divergences associated with the top quark loop are subtracted at
zero momentum:
\begin{equation}
\delta Z_{g_s}=\frac{\alpha_s}{4\pi}\Gamma(1+\epsilon)
(4\pi)^{\epsilon}\left({n_f\over 3}-{11\over 2}\right){1\over
\epsilon_{UV}}+{\alpha_s\over {12\pi}}C_{\epsilon}{1\over
\epsilon_{UV}}.
\end{equation}
Thus, the renormalized strong coupling constant $\alpha_s$ evolves
with $n_f$ light flavors in this scheme. Finally, for the
renormalization constants of the FCNC couplings $\delta
Z_{\kappa^g_{tq}/\Lambda}$, we adopt the $\rm \overline{MS}$ scheme
and adjust it to cancel the remaining UV divergences exactly:
\begin{equation}
\delta
Z_{\kappa^g_{tq}/\Lambda}=\frac{\alpha_s}{6\pi}\Gamma(1+\epsilon)(4\pi)
^{\epsilon}{1\over \epsilon_{UV}},
\end{equation}
and the running of the FCNC couplings are given
by~\cite{Zhang:2009jj}
\begin{equation}
\frac{\kappa^g_{tq}(\mu_r)}{\Lambda}=\frac{\kappa^g_{tq}(m)}{\Lambda}
\left(\frac{\alpha_s(m)}{\alpha_s(\mu_r)}\right)^{2/ (3\beta_0)},
\end{equation}
with $\beta_0=11-2n_f/3$.

The squared amplitudes of the virtual corrections are
\begin{equation}\label{eq5}
\overline{|M|^2}_{gq(qq)}|_{1-loop}=\sum_i2Re\overline{(M^{loop,i}M^{B*})}
_{gq(qq)}+2Re\overline{(M^{con}M^{B*})}_{gq(qq)},
\end{equation}
where $M^{loop,i}_{gq(qq)}$ denote the amplitudes for the i-th loop
diagram in Figs.~\ref{f2}-\ref{f3} or Fig.~\ref{f4}, and $M^{con}_{gq(qq)}$
are the corresponding counterterms. All the one-loop integrals in the loop
amplitudes can be calculated using the standard Passarino-Veltman
techniques~\cite{Denner:1991kt}, and the explicit expressions for the
scalar integrals containing IR divergences can be found in
Ref.~\cite{Ellis:2007qk}. In Eq.~(\ref{eq5}), all the UV divergences
cancel each other, leaving the remaining IR divergences and the finite
terms. Because of the limited space, we do not shown the lengthy explicit
expressions of the virtual corrections here.
The IR divergence of the virtual corrections
can be written as
\begin{eqnarray}\label{eq2}
\overline{|M|^2}_{gq}|_{1-loop,IR}&=&-\frac{\alpha_s}{3\pi}D_{\epsilon}
\bigg\{\frac{11}{\epsilon_{IR}^2}\times\overline{|M^B|^2}_{gq} +{\rm
\ terms\ proportional\ to}\ \frac{1}{\epsilon_{IR}}
\bigg\}, \nonumber \\
\overline{|M|^2}_{qq}|_{1-loop,IR}&=&-\frac{\alpha_s}{\pi}D_{\epsilon}
\bigg\{\frac{2}{\epsilon_{IR}^2}\times\overline{|M^B|^2}_{qq} +{\rm
\ terms\ proportional\ to}\ \frac{1}{\epsilon_{IR}} \bigg\},
\end{eqnarray}
where $D_{\epsilon}=[(4\pi\mu_r^2)/s]^
{\epsilon}/\Gamma(1-\epsilon)$, and $\overline{|M^B|^2}$ are the
squared Born amplitudes given in Eq.~(\ref{eq1}). The
$1/\epsilon_{IR}$ terms can not be factorized in a trivial way due
to the nontrivial color structures of the LO amplitudes, and can be
expressed as combinations of the LO color correlated squared
amplitudes as shown in the next section.

\subsection{Real corrections}\label{ss1}

\begin{figure}[h]
\begin{center}
\includegraphics[width=0.9\textwidth]{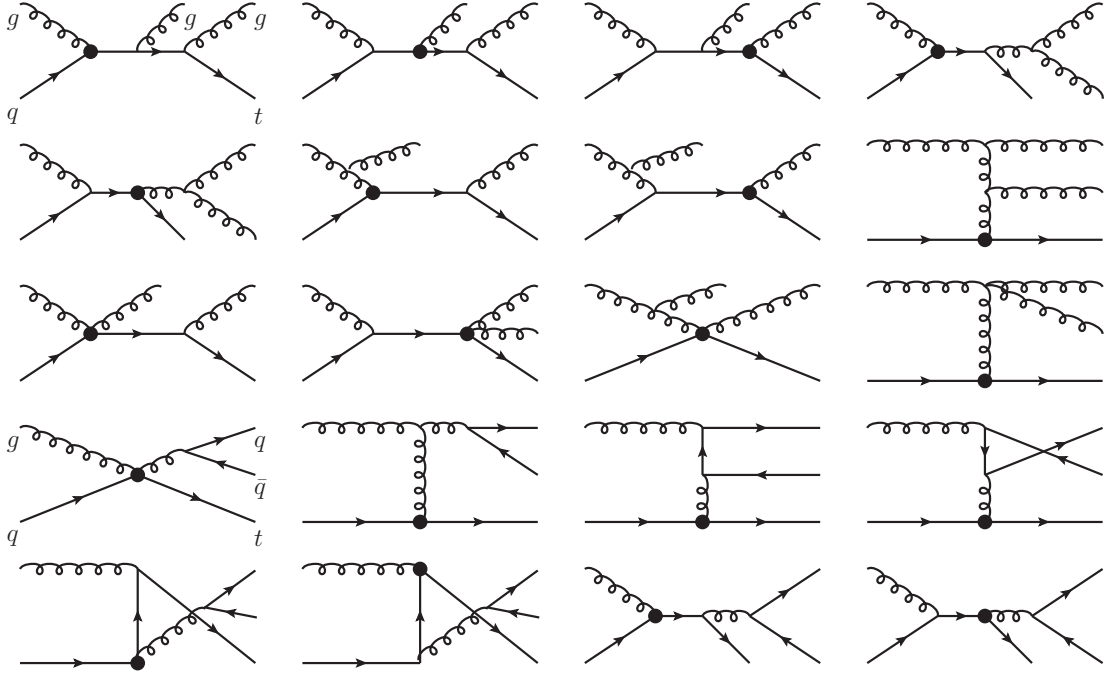}
\caption[]{Typical Feynman diagrams of the real corrections, others
not shown can be obtained by the exchange of the identical external
particles.} \label{f5}
\end{center}
\end{figure}

At the NLO the real corrections consist of the radiations of an
additional gluon or massless (anti)quark in the final states,
including the subprocesses
\begin{eqnarray}
g\ q\longrightarrow t\ g\ g,\ t\ q\ \bar q,\ t\ q'\ \bar{q}',\ g\
g\longrightarrow t\ \bar q\ g,\ q(\bar{q},q')\ q \longrightarrow t \
q(\bar{q},q')\ g,
\end{eqnarray}
as shown in Fig.~\ref{f5}. It should be noted that in our NLO
calculations we did not include the contributions from the SM
on-shell production of the top pair with subsequent rare decay of
one top quark, $pp(\bar p) \to t\bar t\to t +\bar q+ g$, which
provide the same signature as the single top production via the FCNC
couplings and can be calculated separately. We will discuss these
contributions in the numerical results.

Before performing the numerical calculations, we need to extract the
IR divergences in the real corrections. In the dipole formalism this
is done by subtracting some dipole terms from the real corrections
to cancel the singularities and large logarithms exactly, and then
the real corrections become integrable in four dimensions. On the
other hand, these dipole subtraction terms are analytically
integrable in $n$ dimensions over one-parton subspaces, which give
$\epsilon$ poles that represent the soft and collinear divergences.
Then we can add them to the virtual corrections to cancel the
$\epsilon$ poles, and ensure the virtual corrections are also integrable
in four dimensions. This whole procedure can be illustrated by the
formula~\cite{Catani:2002hc}:
\begin{equation}
\hat{\sigma}^{NLO}=\int_{m+1}\left[\left(d\hat{\sigma}^R
\right)_{\epsilon=0}-\left(d\hat{\sigma}^A\right)_{\epsilon=0}\right]
+\int_m\left[d\hat{\sigma}^V+\int_1d\hat{\sigma}^A\right]_{\epsilon=0},
\end{equation}
where $m$ is the number of final state particles at the LO, and
$d\hat{\sigma}^A$ is a sum of the dipole terms. Besides, at hadron
colliders, we have to include the well-known collinear subtraction
counterterms in order to cancel the collinear divergences arising
from the splitting processes of the initial state massless partons.
Here we use the $\overline{\rm MS}$ scheme and the corresponding NLO
PDFs.

For the process with two initial state hadrons, the dipole terms can
be classified into four groups, the final-state emitter and
final-state spectator type,
\begin{eqnarray}
&&\mathcal {D}_{ij,k}(p_1,...,p_{m+1})=\nonumber\\
&&\qquad -{1\over (p_i+p_j)^2-m_{ij}^2}{\
_{m}}\langle...,\widetilde{ij},...,\widetilde
k,...|\frac{\textbf{T}_k\cdot\textbf{T}_{ij}}{\textbf{T}_{ij}^2}
{\textbf{V}_{ij,k}}|..., \widetilde{ij},...,\widetilde
k,...\rangle_{m},
\end{eqnarray}
the final-state emitter and initial-state spectator type,
\begin{eqnarray}
&&\mathcal {D}_{ij}^a(p_1,...,p_{m+1};p_a,...)=\nonumber\\
&&\qquad -{1\over (p_i+p_j)^2-m_{ij}^2}\frac{1}{x_{ij,a}} {\
_{m,a}}\langle...,\widetilde{ij},...;\widetilde
a,...|\frac{\textbf{T}_a\cdot\textbf{T}_{ij}}{\textbf{T}_{ij}^2}
{\textbf{V}_{ij}^a}|..., \widetilde{ij},...;\widetilde
a,...\rangle_{m,a},
\end{eqnarray}
the initial-state emitter and final-state spectator type,
\begin{eqnarray}
&&\mathcal {D}_{j}^{ai}(p_1,...,p_{m+1};p_a,...)=\nonumber\\
&&\qquad -{1\over 2p_a p_i}\frac{1}{x_{ij,a}} {\
_{m,\widetilde{ai}}}\langle...,\widetilde{j},...;\widetilde
{ai},...|\frac{\textbf{T}_j\cdot\textbf{T}_{ai}}{\textbf{T}_{ai}^2}
{\textbf{V}_{j}^{ai}}|..., \widetilde{j},...;\widetilde
{ai},...\rangle_{m,\widetilde{ai}},
\end{eqnarray}
and the initial-state emitter and initial-state spectator type,
\begin{eqnarray}
&&\mathcal {D}^{ai,b}(p_1,...,p_{m+1};p_a,p_b)=\nonumber\\
&&\qquad -{1\over 2p_a p_i}\frac{1}{x_{i,ab}} {\
_{m,\widetilde{ai}}}\langle...;\widetilde
{ai},b|\frac{\textbf{T}_b\cdot\textbf{T}_{ai}}{\textbf{T}_{ai}^2}
{\textbf{V}^{ai,b}}|...;\widetilde {ai},b\rangle_{m,\widetilde{ai}},
\end{eqnarray}
where $a,b$ and $i,j,...$ are the initial and final state partons,
and $\textbf{T}$ and $\textbf{V}$ are the color charge operators and
dipole functions acting on the LO amplitudes, respectively. The
explicit expressions for $x_{i,ab}$, $x_{ij,a}$ and $\textbf{V}$ can
be found in Ref.~\cite{Catani:2002hc}. The integrated dipole
functions together with the collinear counterterms can be written in
the following factorized form
\begin{eqnarray}\label{eq4}
\sim &&\int d\Phi^{(m)}(p_a,p_b) \
_{m,ab}\langle...;p_a,p_b|\textbf{I}_{m+a+b}(\epsilon)|
...;p_a,p_b\rangle_{m,ab}\nonumber\\
&&+\sum_{a'}\int_0^1 dx\int d\Phi^{(m)}(xp_a,p_b) _{m,a'b}\langle
...;xp_a,p_b|\textbf{P}_{m+b}^{a,a'}
(x)+\textbf{K}_{m+b}^{a,a'}(x)|...;xp_a,p_b\rangle_{m,a'b}
\nonumber\\
&&\qquad \qquad \qquad +(a\leftrightarrow b),
\end{eqnarray}
where $x$ is the momentum fraction of the splitting parton,
$d\Phi^{(m)}$ contains all the factors apart from the squared
amplitudes, $\textbf{I}$, $\textbf{P}$, and $\textbf{K}$ are
insertion operators defined in~\cite{Catani:2002hc}. For simplicity,
in all the above formulas we do not show the jet functions that
define the observables and are included in our numerical
calculations.

The operators $\textbf{P}$ and $\textbf{K}$ provide finite
contributions to the NLO corrections, and only the operator
$\textbf{I}$ contains the IR divergences
\begin{eqnarray}\label{eq3}
\textbf{I}|_{IR}&=&-\frac{\alpha_s}{2\pi}{(4\pi)^{\epsilon}\over
\Gamma(1-\epsilon)}\bigg\{\sum_j\sum_{k\neq j}\textbf{T}_j\cdot
\textbf{T}_k\bigg[\left(\frac{\mu_r^2}{s_{jk}}\right)^{\epsilon}
\mathcal{V}(s_{jk},m_j,m_k;\epsilon_{IR})
+{1\over \textbf{T}_j^2}\Gamma_j(m_j,\epsilon_{IR})\bigg]\nonumber\\
&& +\sum_j\textbf{T}_j\cdot\textbf{T}_a\bigg[2\left(
\frac{\mu_r^2}{s_{ja}}\right)^{\epsilon}\mathcal{V}(s_{ja},m_j,0;\epsilon_{IR})
+{1\over\textbf{T}_j^2}\Gamma_j(m_j,\epsilon_{IR})
+{1\over\textbf{T}_a^2}\frac{\gamma_a}{\epsilon_{IR}}\bigg]
\nonumber\\&&
+\textbf{T}_a\cdot\textbf{T}_b\bigg[\left(
\frac{\mu_r^2}{s_{ab}}\right)^{\epsilon}\left(\frac{1}
{\epsilon_{IR}^2}+{1\over\textbf{T}_a^2}\frac{\gamma_a}
{\epsilon_{IR}}\right)\bigg]+(a\leftrightarrow b)\bigg\},
\end{eqnarray}
with
\begin{eqnarray}
\mathcal{V}(s_{jk},m_j,m_k;\epsilon_{IR})&=&{1\over
v_{jk}}\left(\frac{Q_{jk}^2}{s_{jk}}\right)^{\epsilon}\times \left(
1-{1\over 2}\rho_j^{-2\epsilon}-{1\over 2}\rho_k^{-2\epsilon}\right)
{1\over \epsilon_{IR}^2},\nonumber\\
\Gamma_j(0,\epsilon_{IR})&=&\frac{\gamma_j}{\epsilon_{IR}},\quad
\Gamma_j(m_j\neq 0,\epsilon_{IR})=\frac{C_F}{\epsilon_{IR}},
\end{eqnarray}
where $C_F=4/3$, $\gamma_q=2$, and $\gamma_g=11/2-n_f/3$. And
$s_{jk}$, $Q_{jk}^2$, $v_{jk}$, and $\rho_n$ are kinematic variables
defined as follows
\begin{eqnarray}
s_{jk}&=&2p_jp_k,\quad Q_{jk}^2=s_{jk}+m_j^2+m_k^2,\quad
v_{jk}=\sqrt{1-\frac{m_j^2m_k^2}{(p_jp_k)^2}}, \nonumber\\
\rho_n&=&\sqrt{\frac{1-v_{jk}+2m_n^2/(Q_{jk}^2-m_j^2-m_k^2)}
{1+v_{jk}+2m_n^2/(Q_{jk}^2-m_j^2-m_k^2)}}\quad (n=j,k).
\end{eqnarray}
When inserting Eq.~(\ref{eq3}) into the LO amplitudes for the $gq$
and $qq$ subprocesses as shown in Eq.~(\ref{eq4}), we can see that
the IR divergences, including the $1/\epsilon_{\rm IR}$ terms, can
be written as combinations of the LO color correlated squared
amplitudes and all the IR divergences from the virtual corrections
in Eq.~(\ref{eq2}) are canceled exactly, as we expected.

\section{Numerical Results}\label{s4}

In the numerical calculations, we investigate the NLO QCD effects on
the total cross sections, the scale dependence, and several
important distributions at both the Tevatron and LHC. For the single
top production via the FCNC couplings at the NLO, the final state
consists of a top quark plus one or two partons which may form jets.
We use the cone jet algorithm with $\Delta R=0.5$, and require the
jet to have $p_T>20{\rm GeV}$ and $|\eta|<2.5$. We require that
there is at least one jet in the final state, which means the signal
we considered is $t+jet+X$, unlike the direct top production.
All the input parameters are taken
to be~\cite{Amsler:2008zzb}:
\begin{equation}
m_t=171.2{\rm TeV},\quad \alpha_s(M_Z)=0.118,\quad
\kappa^g_{tu}/\Lambda=\kappa^g_{tc}/\Lambda=0.01{\rm TeV}^{-1}.
\end{equation}
The running QCD coupling constant is evaluated at the three-loop
order~\cite{Amsler:2008zzb} and the CTEQ6M PDF
set~\cite{Pumplin:2002vw} is used throughout the calculations of the
NLO (LO) cross sections. Both the renormalization and factorization
scales are fixed to the top quark mass unless specified. We have performed
two independent calculations for the virtual corrections and the integrated
dipole terms, and used the modified MadDipole~\cite{Frederix:2008hu}
package to generate the Fortran code for the real corrections.
The numerical results of the two groups are in good agreement
within the expected accuracy of our numerical program.

\begin{table}[h]
\begin{center}
\begin{tabular}{ccccc}
  \hline
  \hline
  FCNC coupling & $tug$  (LO) & $tug$  (NLO) & $tcg$  (LO) & $tcg$  (NLO) \\
  \hline
  LHC $(\frac{\kappa/\Lambda}{0.01{\rm TeV}^{-1}})^2$ pb & 6.77 & 8.41 & 1.10 & 1.49 \\
  \hline
  Tevatron $(\frac{\kappa/\Lambda}{0.01{\rm TeV}^{-1}})^2$ fb & 86  & 129  & 6.2 & 10.2 \\
  \hline
  \hline
\end{tabular}
\end{center}
\caption{The LO and NLO total cross sections for the single top
quark production via the FCNC couplings at both the LHC and
Tevatron.} \label{t1}
\end{table}

\begin{figure}[H]
  \subfigure{
    \begin{minipage}[b]{0.5\textwidth}
      \begin{center}
     \scalebox{0.4}{\includegraphics*[50,20][580,420]{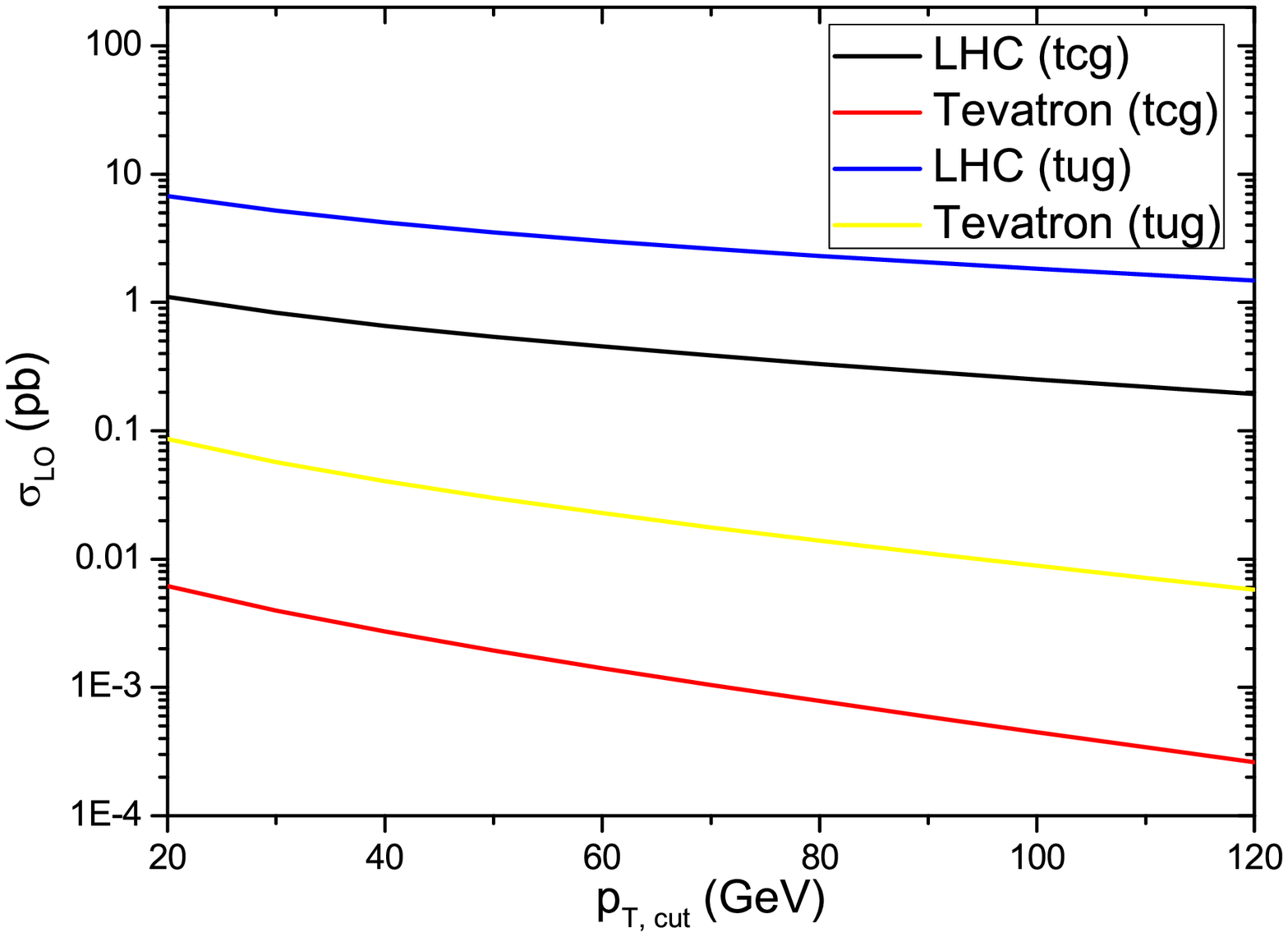}}
      \end{center}
    \end{minipage}}
  \subfigure{
    \begin{minipage}[b]{0.5\textwidth}
      \begin{center}
     \scalebox{0.4}{\includegraphics*[50,20][580,420]{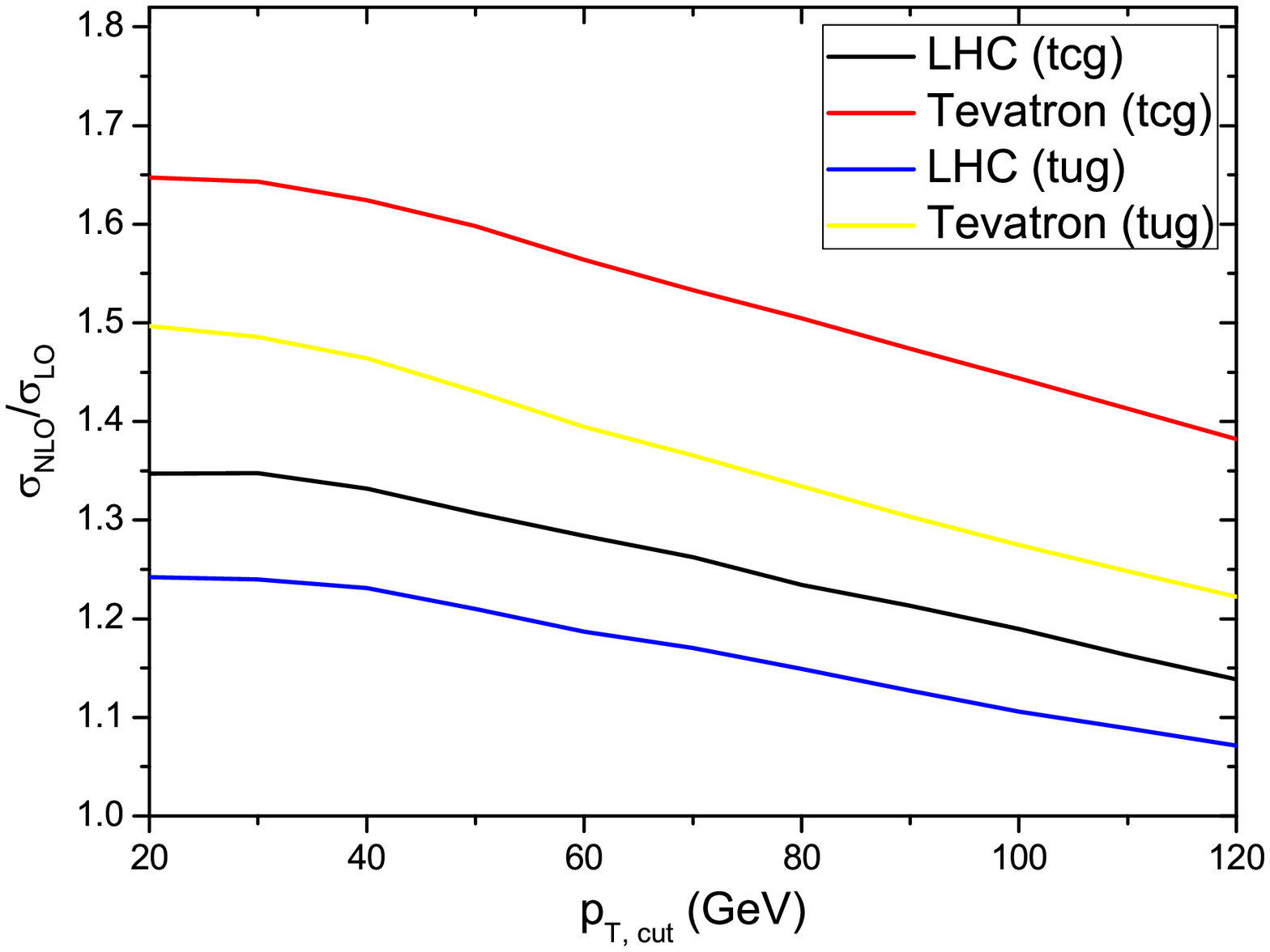}}
      \end{center}
    \end{minipage}}
    \caption{\label{f6}The LO total cross sections and NLO K factors
    as functions of the leading jet transverse momentum cut.}

\end{figure}

In Table~\ref{t1}, we list some typical numerical results of the LO
and NLO total cross sections for the single top quark production via
the FCNC couplings. We can see that the cross sections can reach
about 8 pb and 1 pb at the LHC, for the $tug$ and $tcg$ coupling,
respectively. But if we take the FCNC coupling values to the
experimental upper limits~\cite{Aaltonen:2008qr}, then the cross
sections can reach as large as several tens of pb. In Fig.~\ref{f6}
we present the LO total cross sections and the K factors
$\sigma_{NLO}/\sigma_{LO}$ as functions of the leading jet
transverse momentum cut, respectively. It can be seen that, for the
$tcg$ coupling the NLO corrections can enhance the total cross
sections by about 60\% and 30\%, and for the $tug$ coupling by about
50\% and 20\% at the Tevatron and LHC, respectively. And the K
factors decrease with the increasing transverse momentum cut. In
Sec.~\ref{ss1} we mentioned that the SM on shell production of the
top pair with subsequent decay also contributes to the same final
state as the FCNC single top production. According to
Ref.~\cite{Zhang:2008yn}, the former contributions are about
$7\times10^{-4}$ pb and $8\times10^{-2}$ pb at the Tevatron and LHC,
respectively, for our chosen FCNC coupling values. Compared with the
results in Table~\ref{t1}, we can see that these contributions are
negligibly small for the $tug$ coupling, but can reach about 10\% of
the LO total cross sections for the $tcg$ coupling at both the
Tevatron and LHC.

\begin{figure}[H]
  \subfigure{
    \begin{minipage}[b]{0.5\textwidth}
      \begin{center}
     \scalebox{0.4}{\includegraphics*[50,20][570,420]{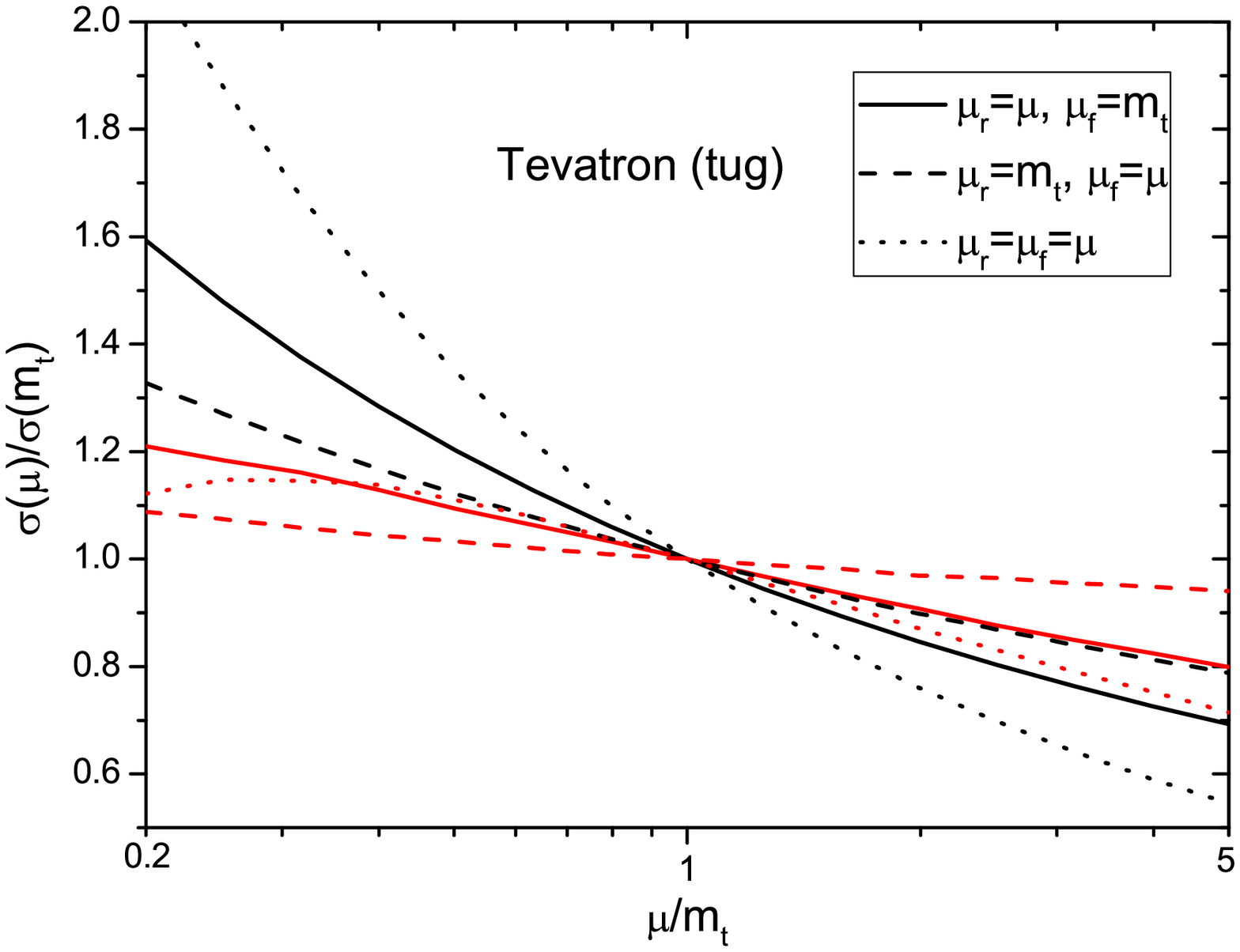}}
      \end{center}
    \end{minipage}}
  \subfigure{
    \begin{minipage}[b]{0.5\textwidth}
      \begin{center}
     \scalebox{0.4}{\includegraphics*[50,20][570,420]{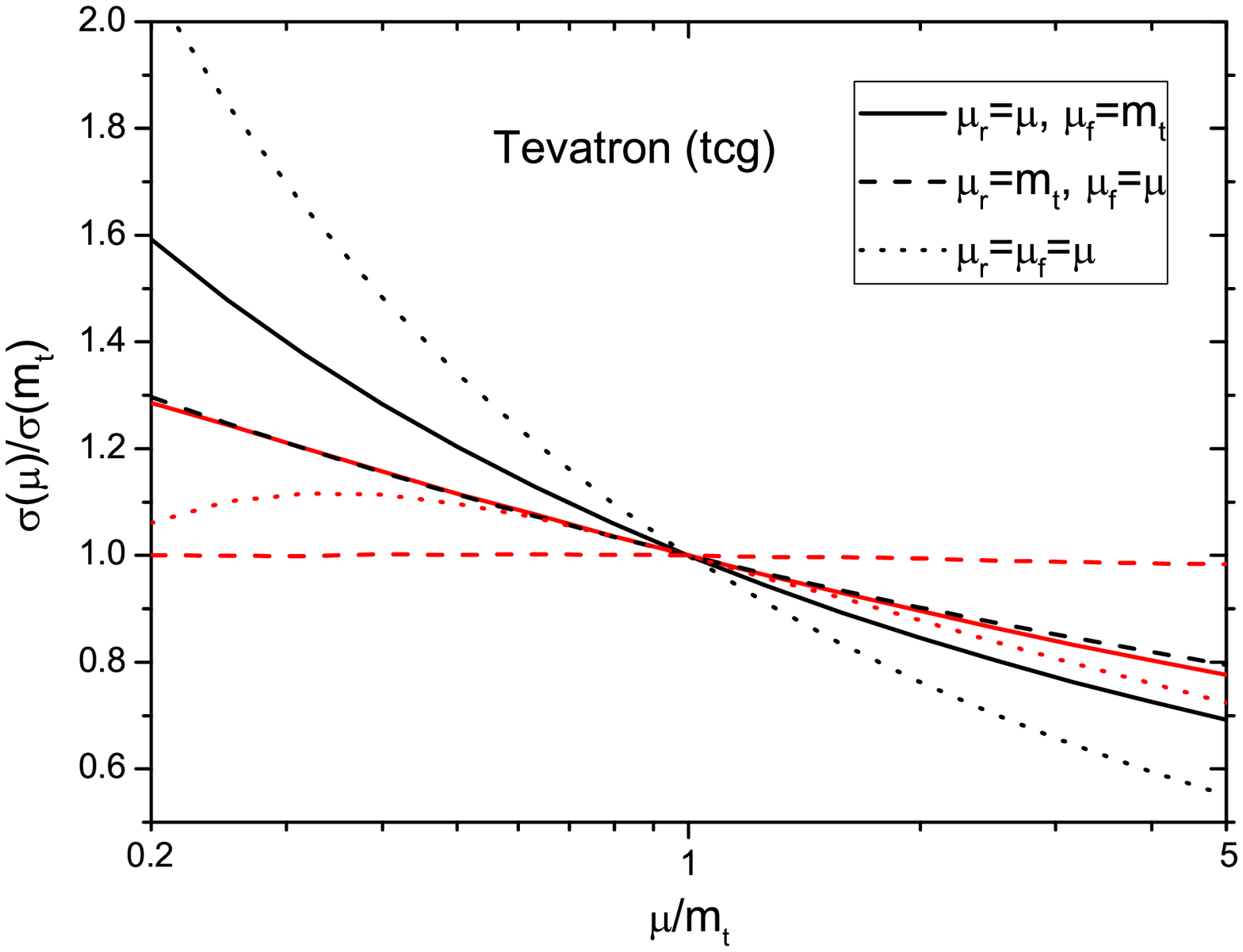}}
      \end{center}
    \end{minipage}}
    \caption{\label{f7}Scale dependence of the total cross sections at the
    Tevatron, the black lines represent the LO results, while the red ones
    represent the NLO results.}
\end{figure}

\begin{figure}[H]
  \subfigure{
    \begin{minipage}[b]{0.5\textwidth}
      \begin{center}
     \scalebox{0.40}{\includegraphics*[50,20][570,420]{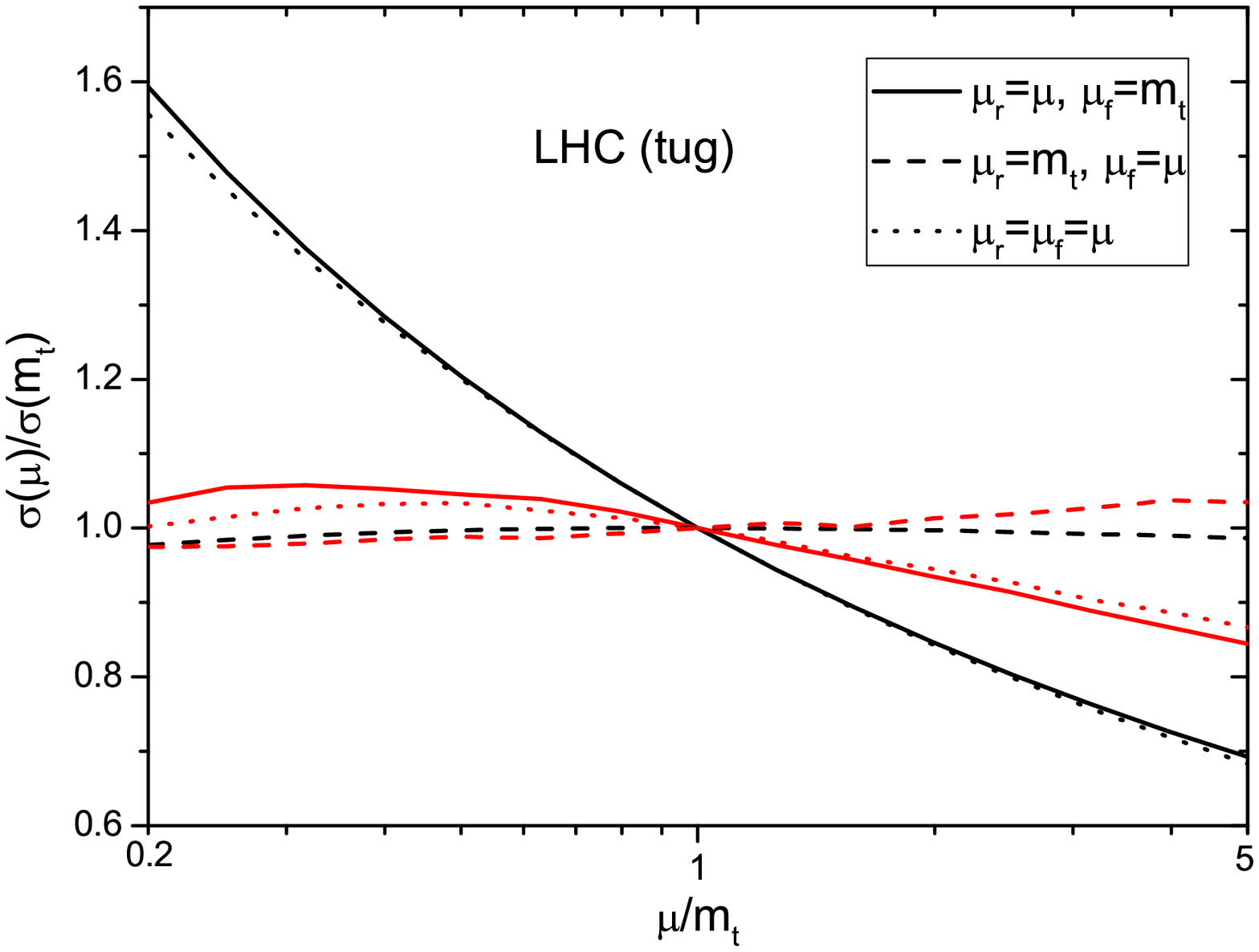}}
      \end{center}
    \end{minipage}}
  \subfigure{
    \begin{minipage}[b]{0.5\textwidth}
      \begin{center}
     \scalebox{0.40}{\includegraphics*[50,20][570,420]{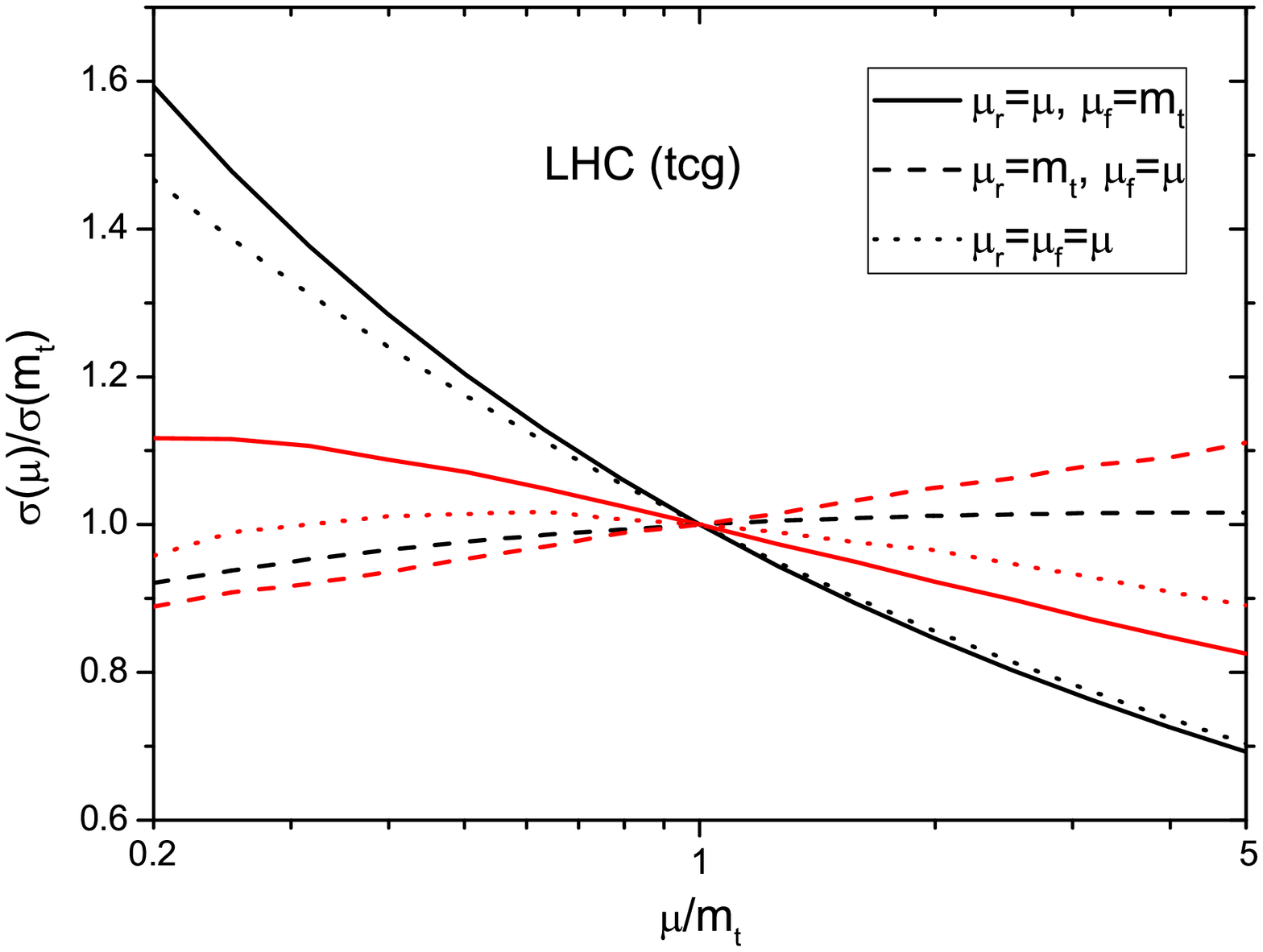}}
      \end{center}
    \end{minipage}}
    \caption{\label{f8}Scale dependence of the total cross sections at the LHC,
    the black lines represent the LO results, while the red ones represent
    the NLO results.}
\end{figure}

\begin{figure}[H]
\begin{center}
\scalebox{0.45}{\includegraphics*[50,20][570,420]{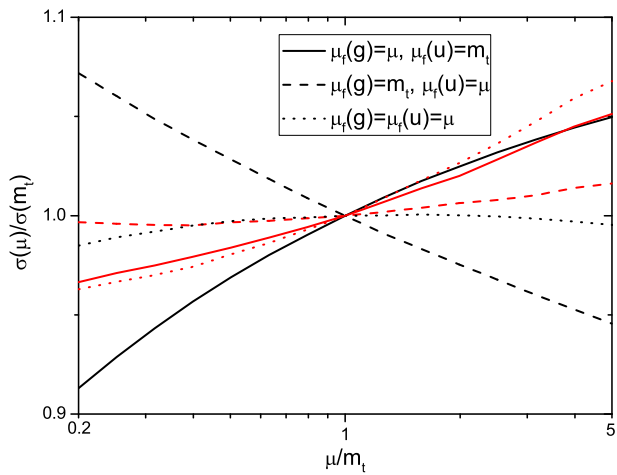}}
\caption[]{Factorization scale dependence for the $gu$ initial state
subprocess at the LHC, the black lines represent the LO results,
while the red ones represent the NLO results.} \label{f14}
\end{center}
\end{figure}

In Figs.~\ref{f7} and ~\ref{f8} we show the scale dependence of the
LO and NLO total cross section for three cases: (1) the
renormalization scale dependence $\mu_r=\mu,\ \mu_f=m_t$, (2) the
factorization scale dependence $\mu_r=m_t,\ \mu_f=\mu$, and (3) total
scale dependence $\mu_r=\mu_f=\mu$. It can be seen that the NLO
corrections reduce the scale dependence significantly for all three
cases, which make the theoretical predictions more reliable. For
example, at the Tevatron for the $tug$ coupling, when the scale
$\mu$ varies from $0.2m_t$ to $5m_t$, the variations of the total
cross sections are 90\% and 40\% for case (1), 50\% and 20\% for
case (2), and 150\% and 50\% for case (3) at the LO and NLO,
respectively. Note that, from Fig.~\ref{f8} it seems that the NLO
corrections did not reduce the factorization scale dependence at the
LHC. This is because the PDFs of the two incoming partons have
opposite trends with the increasing factorization scale and the
scale dependence cancel each other at the LO, but there is no such
cancellation at the NLO. In order to further explain this scenario,
in Fig.~\ref{f14} we plot the factorization scale dependence of the
$gu$ initial state subprocess at the LHC, which contributes about
90\% of the LO total cross section for the $tug$ coupling. It can be
seen that, if we fix the factorization scale of one incoming parton
and only change another one, the NLO corrections indeed improve the
factorization scale dependence.

\begin{figure}[H]
  \subfigure{
    \begin{minipage}[b]{0.5\textwidth}
      \begin{center}
     \scalebox{0.45}{\includegraphics*[40,15][520,380]{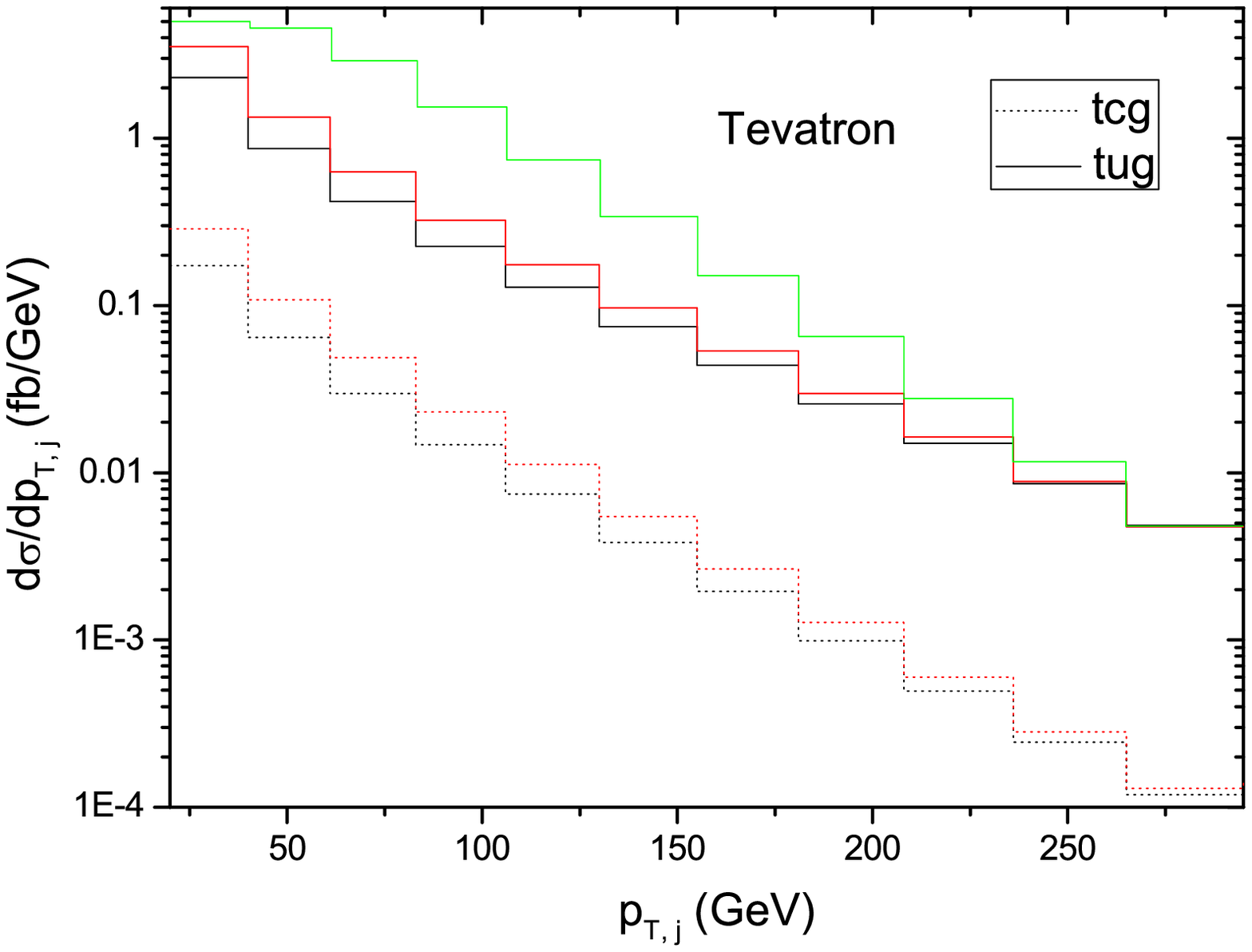}}
      \end{center}
    \end{minipage}}
  \subfigure{
    \begin{minipage}[b]{0.5\textwidth}
      \begin{center}
     \scalebox{0.45}{\includegraphics*[40,15][520,380]{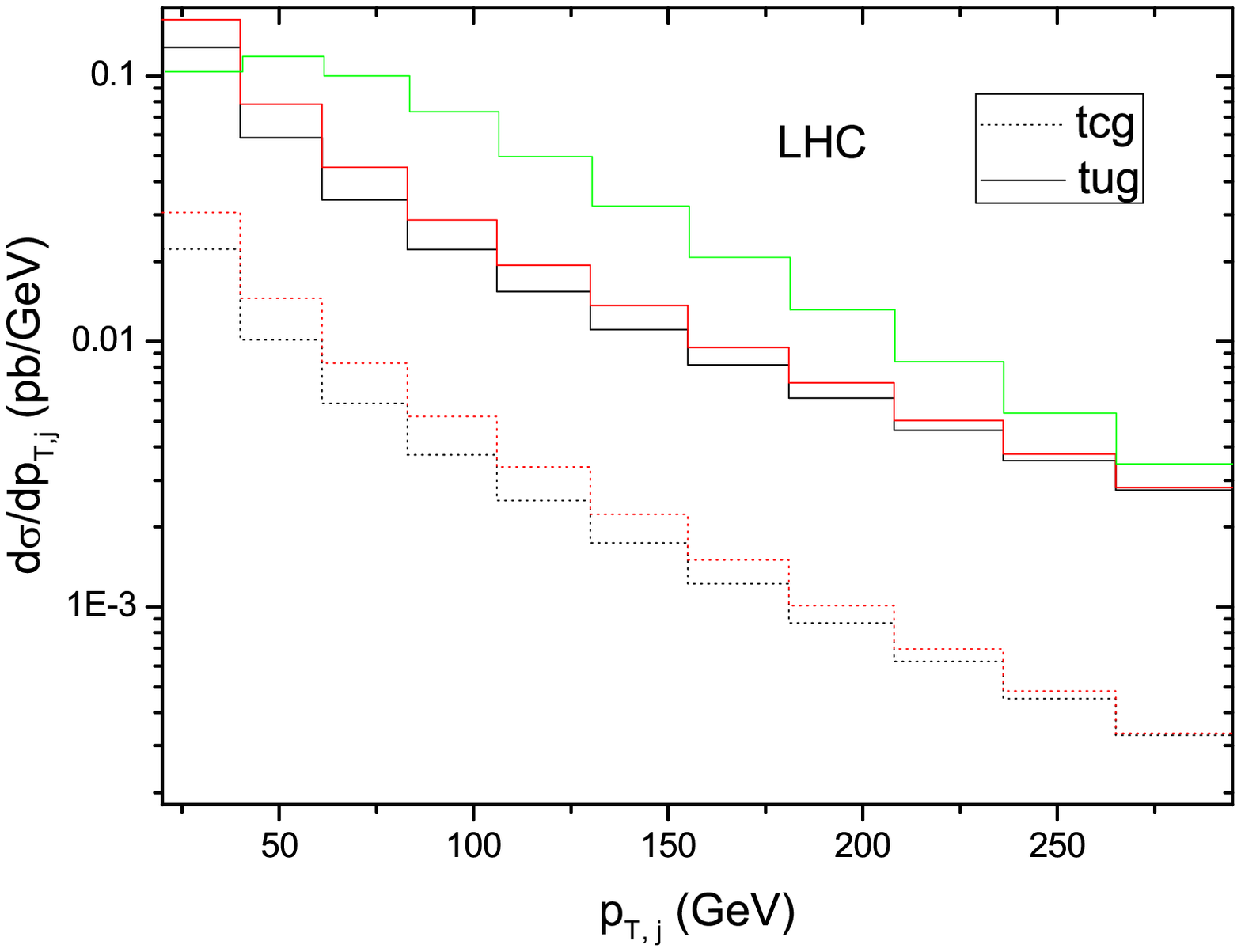}}
      \end{center}
    \end{minipage}}
    \caption{\label{f9}Transverse momentum distributions of the leading jet, the black
    and red lines represent the LO and NLO results of the FCNC single top
    production, respectively, while the green lines correspond to the SM
    ones normalized to arbitrary units.}
\end{figure}

\begin{figure}[H]
  \subfigure{
    \begin{minipage}[b]{0.5\textwidth}
      \begin{center}
     \scalebox{0.45}{\includegraphics*[40,15][520,380]{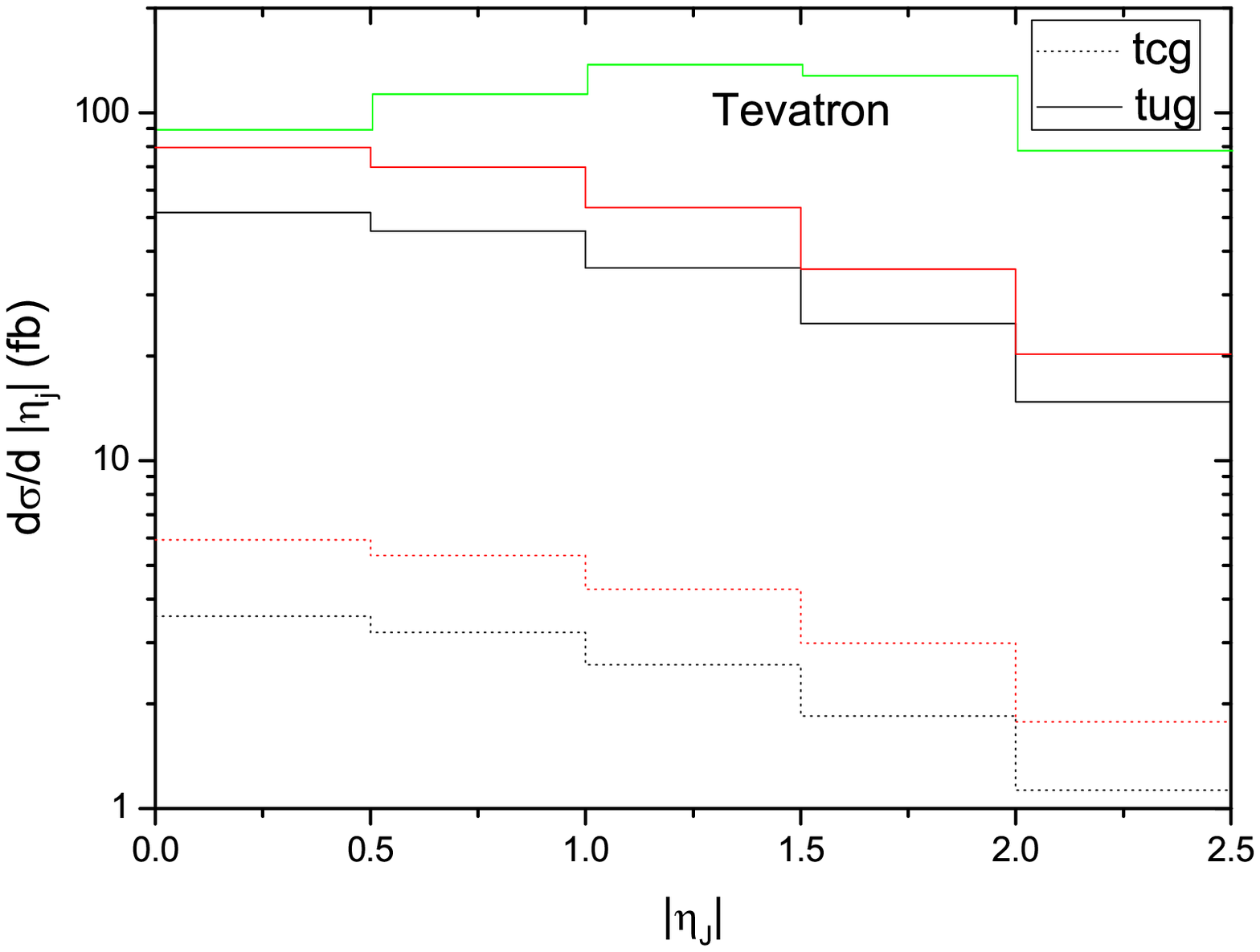}}
      \end{center}
    \end{minipage}}
  \subfigure{
    \begin{minipage}[b]{0.5\textwidth}
      \begin{center}
     \scalebox{0.45}{\includegraphics*[40,15][520,380]{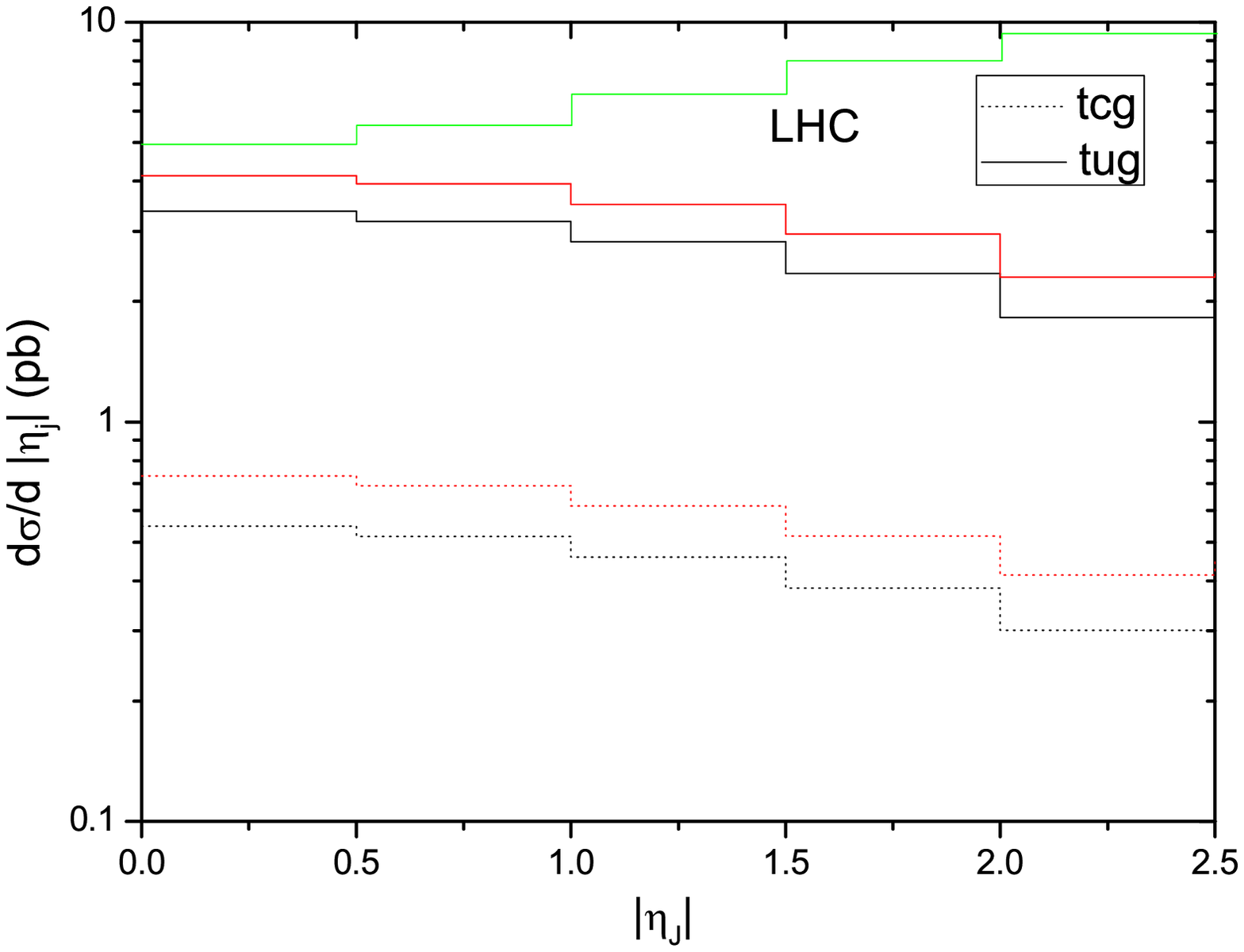}}
      \end{center}
    \end{minipage}}
    \caption{\label{f10}Pseudorapidity distributions of the leading jet, the black
    and red lines represent the LO and NLO results of the FCNC single top
    production, respectively, while the green lines correspond to the SM
    ones normalized to arbitrary units.}
\end{figure}

Figure~\ref{f9} shows the transverse momentum distributions of the
leading jet for the single top production via the FCNC couplings
together with the SM t-channel single top production, which is the
main SM background of our process. We can see that the distributions
of the single top production via the FCNC couplings decrease more
quickly than the SM ones with the increasing $p_T$, and the NLO
corrections increase the distributions of the FCNC single top
production, especially in the low $p_T$ regions. In Fig.~\ref{f10}
we show the pseudorapidity distributions of the leading jet. For the
FCNC single top production, the distributions decrease with the
increasing pseudorapidity at both the Tevatron and LHC, while for
the SM ones the distributions are almost flat at the Tevatron and
increase at the LHC. The NLO corrections increase the distributions
by almost the same amount in all the regions.

\begin{figure}[H]
  \subfigure{
    \begin{minipage}[b]{0.5\textwidth}
      \begin{center}
     \scalebox{0.45}{\includegraphics*[40,15][520,380]{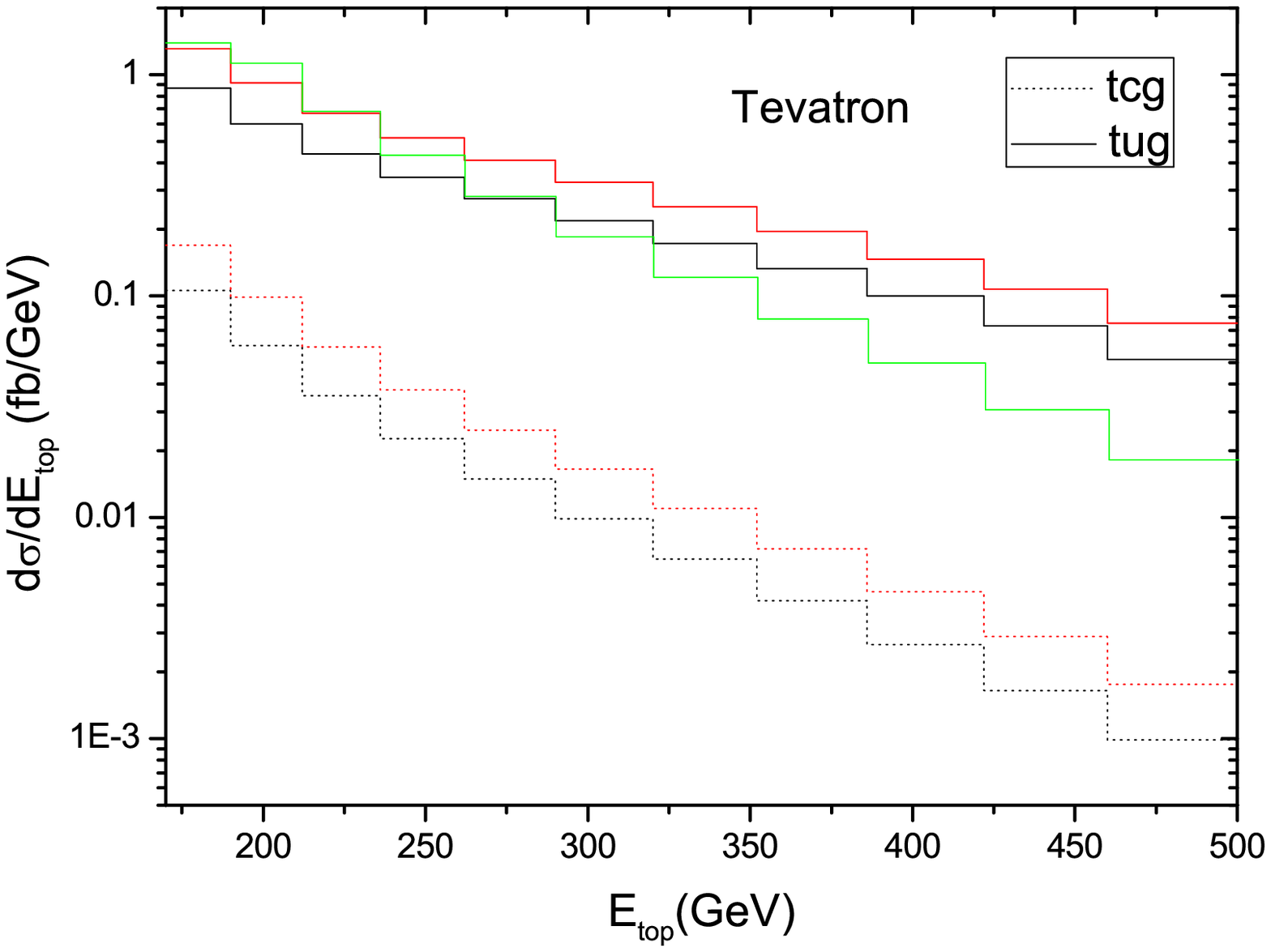}}
      \end{center}
    \end{minipage}}
  \subfigure{
    \begin{minipage}[b]{0.5\textwidth}
      \begin{center}
     \scalebox{0.45}{\includegraphics*[40,15][520,380]{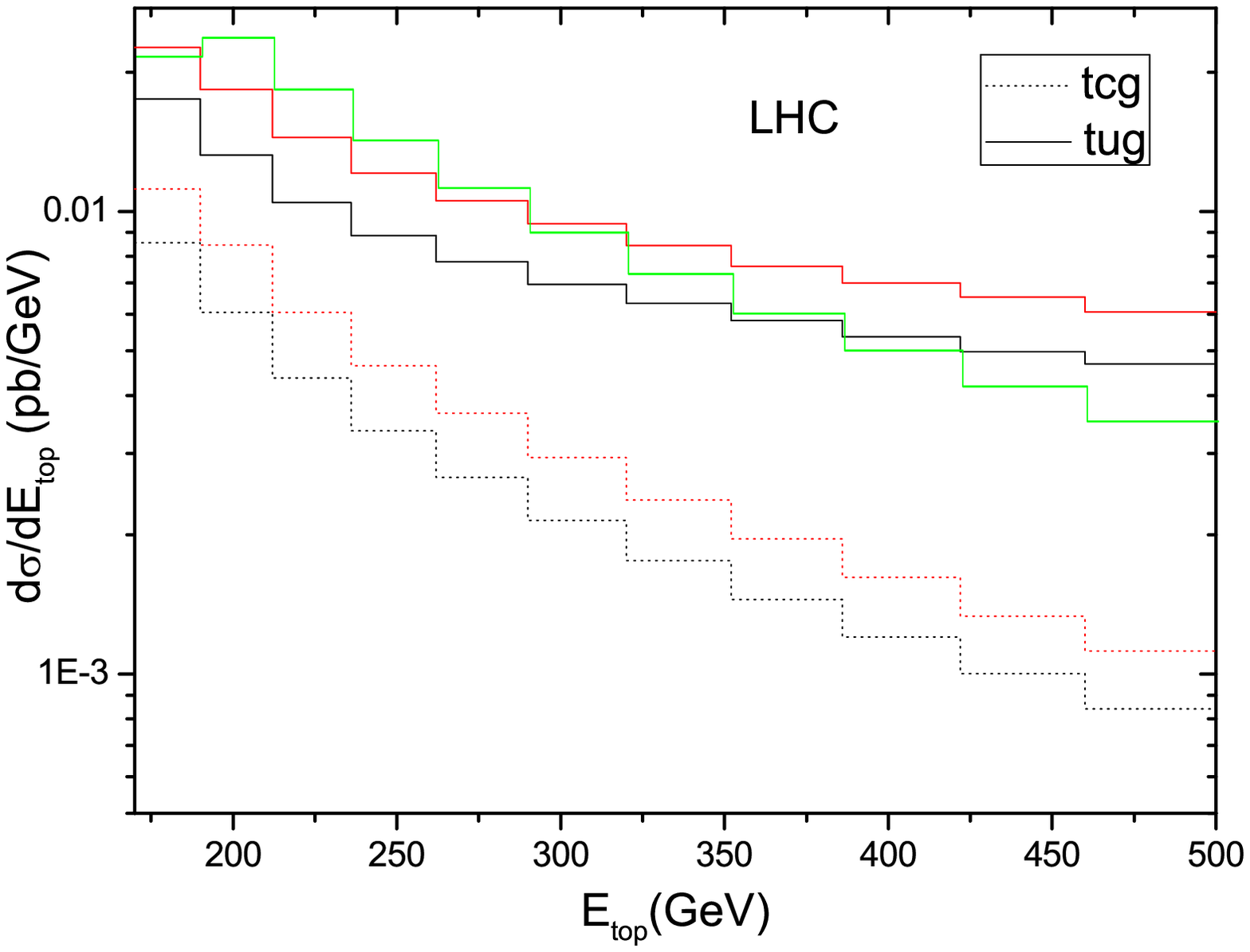}}
      \end{center}
    \end{minipage}}
    \caption{\label{f11}Energy distributions of the top
    quark, the black and red lines represent the LO and NLO results of the FCNC
    single top production, respectively, while the green lines correspond to the SM
    ones normalized to arbitrary units.}
\end{figure}

\begin{figure}[H]
  \subfigure{
    \begin{minipage}[b]{0.5\textwidth}
      \begin{center}
     \scalebox{0.45}{\includegraphics*[40,15][520,380]{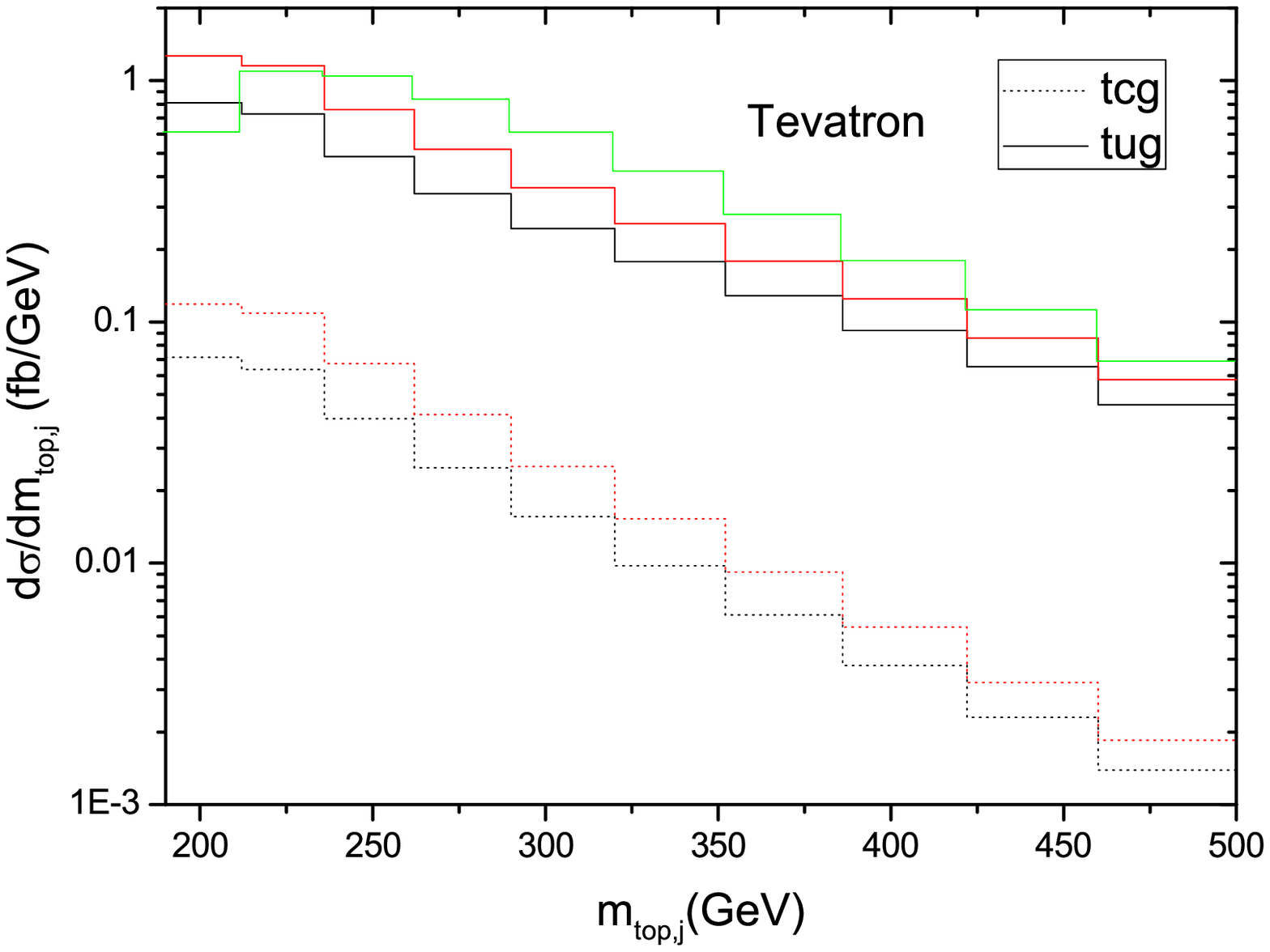}}
      \end{center}
    \end{minipage}}
  \subfigure{
    \begin{minipage}[b]{0.5\textwidth}
      \begin{center}
     \scalebox{0.45}{\includegraphics*[40,15][520,380]{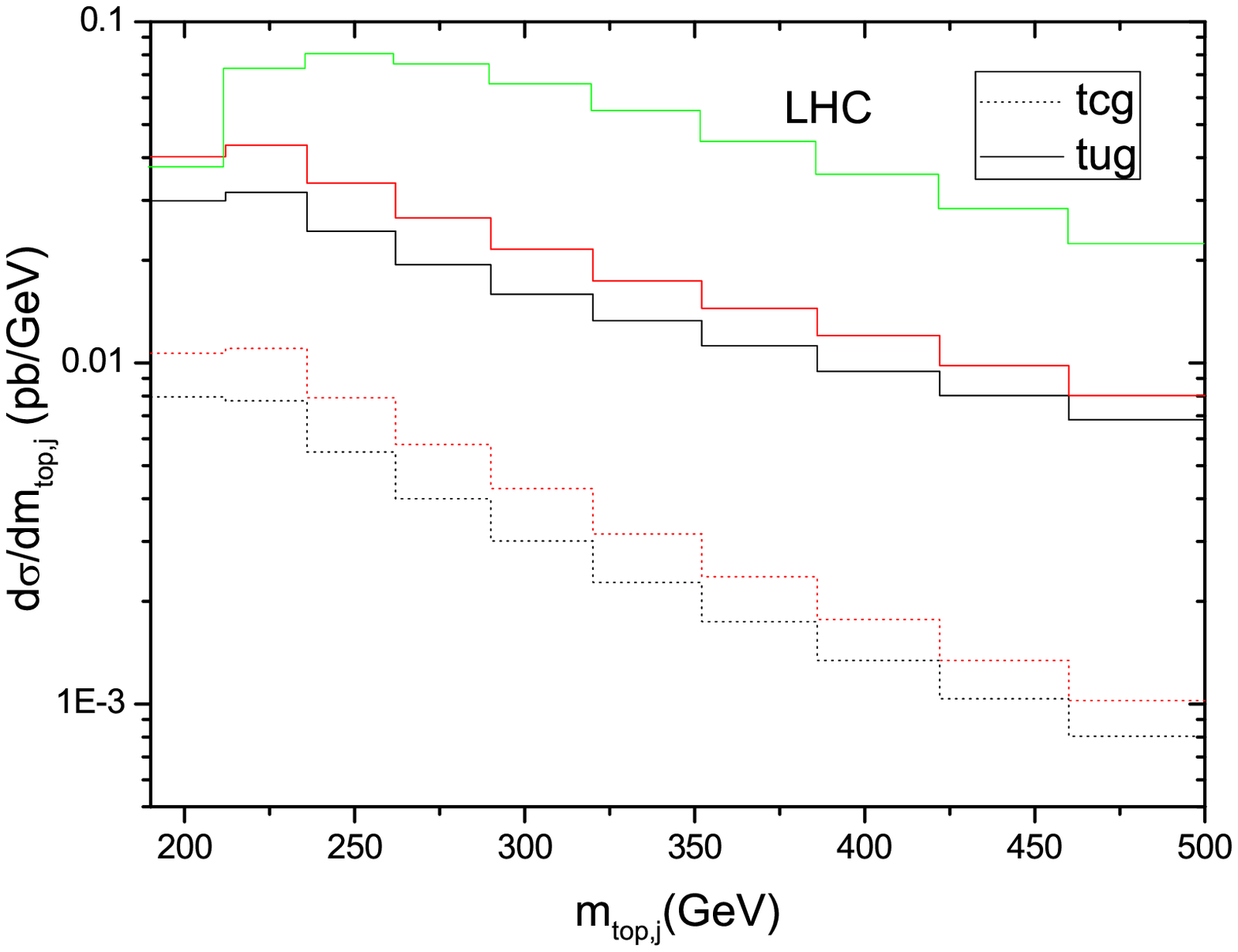}}
      \end{center}
    \end{minipage}}
    \caption{\label{f12}Invariant mass distributions of the leading jet and the top
    quark, the black and red lines represent the LO and NLO results of the FCNC
    single top production, respectively, while the green lines correspond to the SM
    ones normalized to arbitrary units.}
\end{figure}

Fig.~\ref{f11} gives the top quark energy distributions. We can see
that the SM ones decrease faster than the one of $tug$ FCNC coupling
with the increasing top quark energy. Figure~\ref{f12} shows the
invariant mass distributions of the top quark and the leading jet.
The shapes for the FCNC single top production are different from the
SM ones, where there is a peak in the middle region. The NLO
corrections do not change the shapes of these two distributions.

\begin{figure}[H]
  \subfigure{
    \begin{minipage}[b]{0.5\textwidth}
      \begin{center}
     \scalebox{0.45}{\includegraphics*[40,20][520,380]{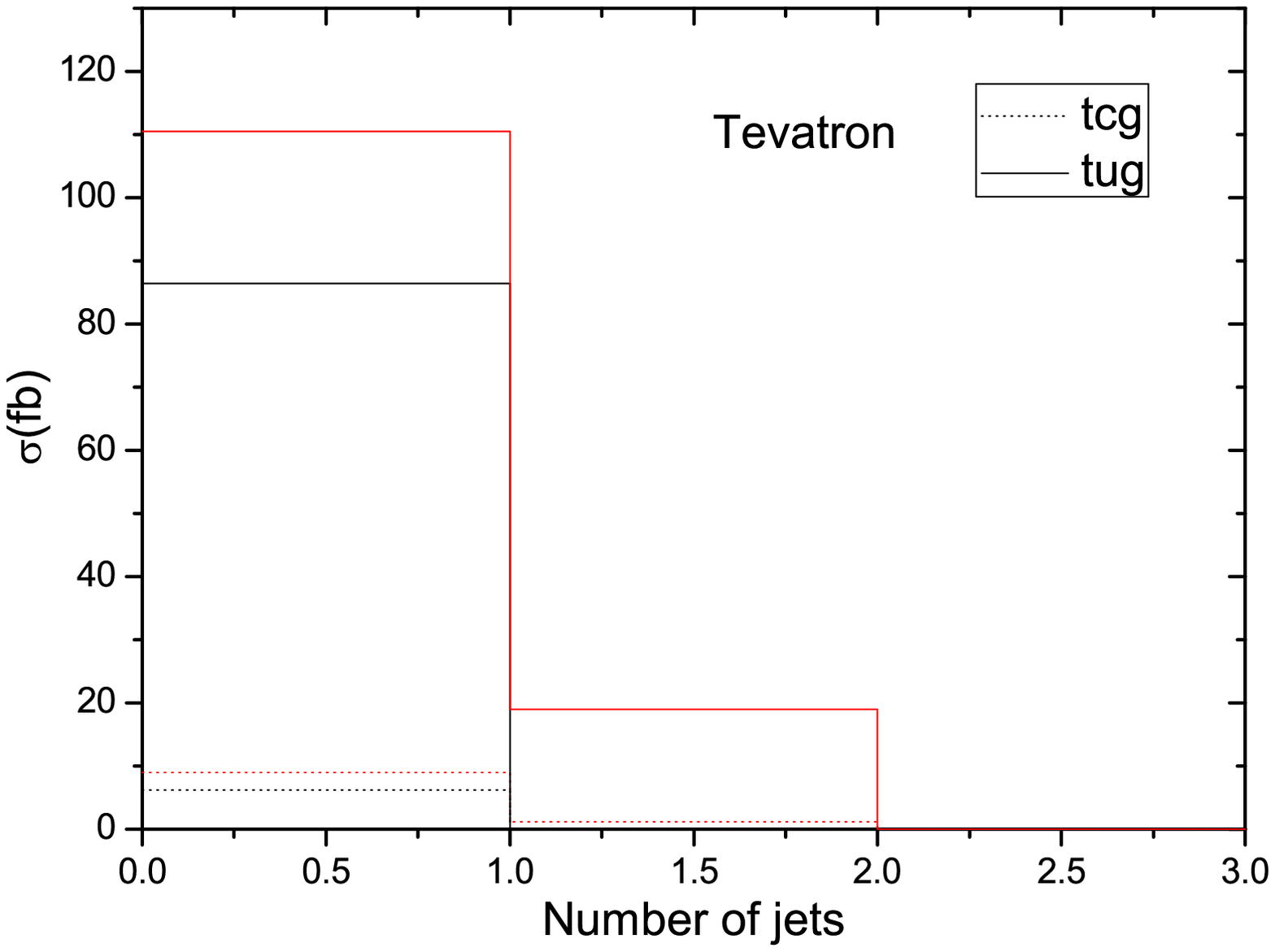}}
      \end{center}
    \end{minipage}}
  \subfigure{
    \begin{minipage}[b]{0.5\textwidth}
      \begin{center}
     \scalebox{0.45}{\includegraphics*[40,20][520,380]{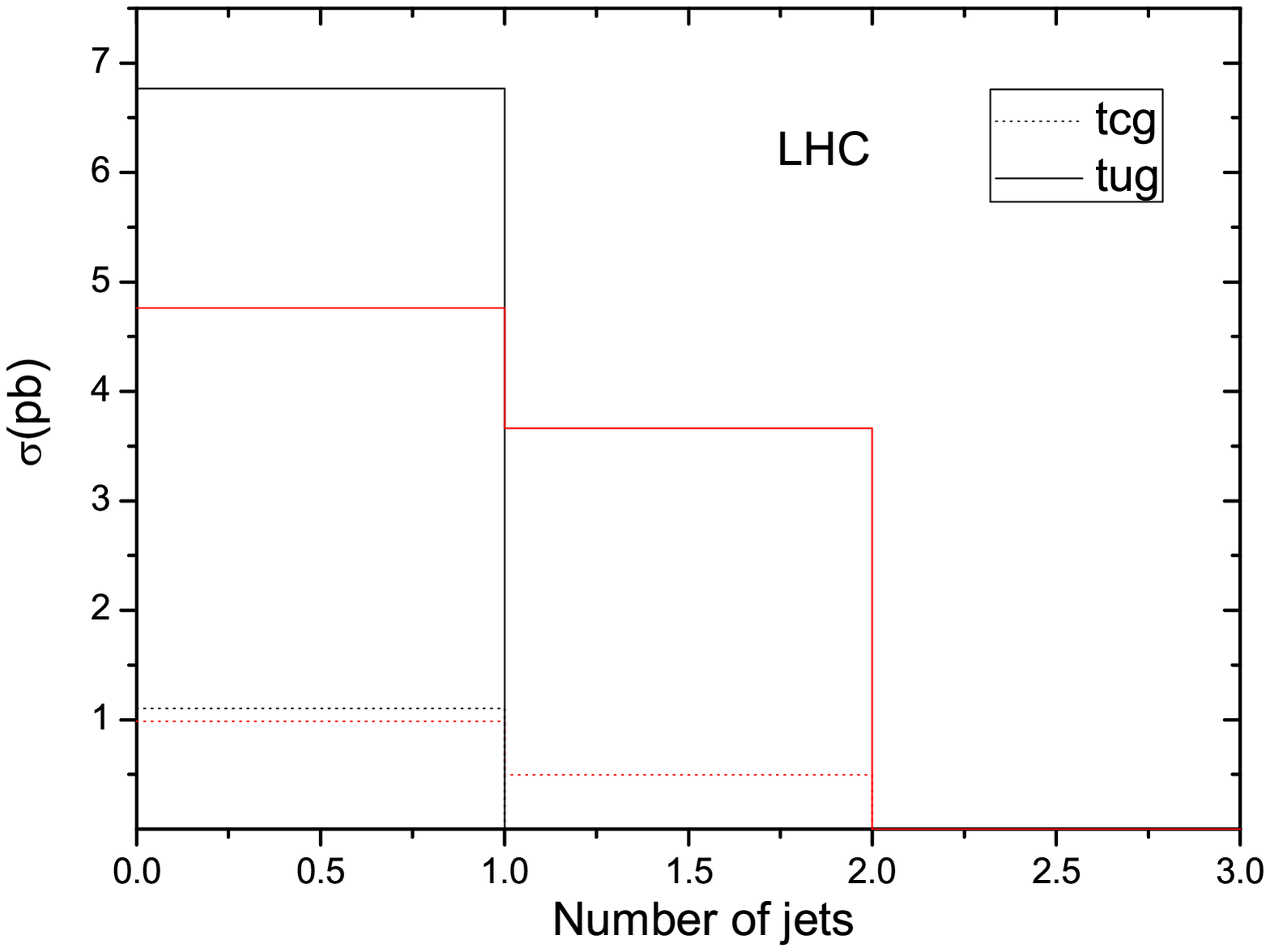}}
      \end{center}
    \end{minipage}}
    \caption{\label{f13}Jet multiplicities distributions of the
    FCNC single top production, the black
    and red lines represent the LO and NLO results, respectively.}
\end{figure}

Figure~\ref{f13} shows the jet multiplicities distributions of the
FCNC single top production. At the LO there is only one jet in the
final state, while at the NLO there may be two jets. For example, at
the LHC, for the $tug$ coupling the LO one jet cross section is 6.7
pb, and the NLO corrections reduce the one jet cross section to 4.8
pb and increase the two jets cross section to 3.7 pb, which is about
50\% of the LO total cross section.

\section{Conclusions}\label{s5}
In conclusion, we have investigated the NLO QCD effects on the
single top productions induced by model-independent $tqg$ FCNC
couplings at both the Tevatron and LHC. Our results show that, for
the $tcg$ coupling the NLO corrections can enhance the total cross
sections by about 60\% and 30\%, and for the $tug$ coupling by about
50\% and 20\% at the Tevatron and LHC, respectively, which means
that the NLO corrections can increase the experimental sensitivity
to the FCNC couplings by about 10\%$-$30\%. Moreover, the NLO
corrections reduce the dependence of the total cross sections on the
renormalization or factorization scale significantly, which lead to
increased confidence on the theoretical predictions. Besides, we
also evaluate the NLO corrections to several important kinematic
distributions, i.e., the transverse momentum and pseudorapidity of
the leading jet, energy of the top quark, jet multiplicities, and
invariant mass of the leading jet and top quark. We find that for
most of them the NLO corrections are almost the same and do not
change the shape of the distributions.

\begin{acknowledgments}
We would like to thank Tao Han for useful
discussions. This work was supported in part by the National Natural
Science Foundation of China, under Grants No. 10721063, No. 10975004
and No. 10635030.
\end{acknowledgments}


\begin{thebibliography}{999}
\bibitem{Glashow:1970gm}
  S.~L.~Glashow, J.~Iliopoulos and L.~Maiani,
  Phys.\ Rev.\  D {\bf 2} (1970) 1285.

\bibitem{AguilarSaavedra:2004wm}
  J.~A.~Aguilar-Saavedra,
  Acta Phys.\ Polon.\  B {\bf 35} (2004) 2695.

\bibitem{delAguila:1998tp}
  F.~del Aguila, J.~A.~Aguilar-Saavedra and R.~Miquel,
  Phys.\ Rev.\ Lett.\  {\bf 82} (1999) 1628;
  J.~A.~Aguilar-Saavedra,
  Phys.\ Rev.\  D {\bf 67} (2003) 035003
  [Erratum-ibid.\  D {\bf 69} (2004) 099901].

\bibitem{Cheng:1987rs}
  T.~P.~Cheng and M.~Sher,
  Phys.\ Rev.\  D {\bf 35} (1987) 3484;
  M.~E.~Luke and M.~J.~Savage,
  Phys.\ Lett.\  B {\bf 307} (1993) 387;
  D.~Atwood, L.~Reina and A.~Soni,
  Phys.\ Rev.\  D {\bf 55} (1997) 3156;
  S.~Bejar, J.~Guasch and J.~Sola,
  Nucl.\ Phys.\  B {\bf 600} (2001) 21.

\bibitem{Li:1993mg}
  C.~S.~Li, R.~J.~Oakes and J.~M.~Yang,
  Phys.\ Rev.\  D {\bf 49} (1994) 293
  [Erratum-ibid.\  D {\bf 56} (1997) 3156];
  J.~L.~Lopez, D.~V.~Nanopoulos and R.~Rangarajan,
  Phys.\ Rev.\  D {\bf 56} (1997) 3100;
  G.~M.~de Divitiis, R.~Petronzio and L.~Silvestrini,
  Nucl.\ Phys.\  B {\bf 504} (1997) 45;
  J.~M.~Yang, B.~L.~Young and X.~Zhang,
  Phys.\ Rev.\  D {\bf 58} (1998) 055001;
  J.~Guasch and J.~Sola,
  Nucl.\ Phys.\  B {\bf 562} (1999) 3;
  G.~Eilam, A.~Gemintern, T.~Han, J.~M.~Yang and X.~Zhang,
  Phys.\ Lett.\  B {\bf 510} (2001) 227;
  J.~j.~Cao, Z.~h.~Xiong and J.~M.~Yang,
  Phys.\ Rev.\ Lett.\  {\bf 88} (2002) 111802;
  J.~J.~Liu, C.~S.~Li, L.~L.~Yang and L.~G.~Jin,
  Phys.\ Lett.\  B {\bf 599} (2004) 92.


\bibitem{Davoudiasl:2001uj}
  H.~Davoudiasl and T.~G.~Rizzo,
  Phys.\ Lett.\  B {\bf 512} (2001) 100;
  P.~M.~Aquino, G.~Burdman and O.~J.~P.~Eboli,
  Phys.\ Rev.\ Lett.\  {\bf 98} (2007) 131601;
  S.~Casagrande, F.~Goertz, U.~Haisch, M.~Neubert and T.~Pfoh,
  JHEP {\bf 0810} (2008) 094;
  J.~Gao, C.~S.~Li, X.~Gao and Z.~Li,
  Phys.\ Rev.\  D {\bf 78} (2008) 096005.


\bibitem{HongSheng:2007ve}
  H.~Hong-Sheng,
  Phys.\ Rev.\  D {\bf 75} (2007) 094010;
  X.~Wang, Y.~Zhang, H.~Jin and Y.~Xi,
  Nucl.\ Phys.\  B {\bf 810} (2009) 226;
  X.~F.~Han, L.~Wang and J.~M.~Yang,
  arXiv:0903.5491 [hep-ph].


\bibitem{Wang:1994qd}
  X.~L.~Wang, G.~R.~Lu, J.~M.~Yang, Z.~J.~Xiao, C.~X.~Yue and Y.~M.~Zhang,
  Phys.\ Rev.\  D {\bf 50} (1994) 5781;
  G.~Burdman,
  Phys.\ Rev.\ Lett.\  {\bf 83} (1999) 2888;
  C.~x.~Yue, G.~r.~Lu, Q.~j.~Xu, G.~l.~Liu and G.~p.~Gao,
  Phys.\ Lett.\  B {\bf 508} (2001) 290;
  J.~j.~Cao, G.~l.~Liu and J.~M.~Yang,
  Phys.\ Rev.\  D {\bf 70} (2004) 114035;
  J.~j.~Cao, G.~l.~Liu, J.~M.~Yang and H.~j.~Zhang,
  Phys.\ Rev.\  D {\bf 76} (2007) 014004.


\bibitem{Buchmuller:1985jz}
  W.~Buchmuller and D.~Wyler,
  Nucl.\ Phys.\  B {\bf 268} (1986) 621.


\bibitem{AguilarSaavedra:2008zc}
  J.~A.~Aguilar-Saavedra,
  Nucl.\ Phys.\  B {\bf 812} (2009) 181.


\bibitem{Hosch:1997gz}
  M.~Hosch, K.~Whisnant and B.~L.~Young,
  Phys.\ Rev.\  D {\bf 56} (1997) 5725.


\bibitem{Malkawi:1995dm}
  E.~Malkawi and T.~M.~P.~Tait,
  Phys.\ Rev.\  D {\bf 54} (1996) 5758;
  P.~M.~Ferreira and R.~Santos,
  Phys.\ Rev.\  D {\bf 73} (2006) 054025.


\bibitem{Han:1998tp}
  T.~Han, M.~Hosch, K.~Whisnant, B.~L.~Young and X.~Zhang,
  Phys.\ Rev.\  D {\bf 58} (1998) 073008.


\bibitem{Carvalho:2007yi}
  J.~Carvalho {\it et al.}  [ATLAS Collaboration],
  Eur.\ Phys.\ J.\  C {\bf 52} (2007) 999.


\bibitem{Aaltonen:2008qr}
  T.~Aaltonen {\it et al.}  [CDF Collaboration],
  Phys.\ Rev.\ Lett.\  {\bf 102} (2009) 151801.


\bibitem{Abazov:2007ev}
  V.~M.~Abazov {\it et al.}  [D0 Collaboration],
  Phys.\ Rev.\ Lett.\  {\bf 99} (2007) 191802.


\bibitem{Liu:2005dp}
  J.~J.~Liu, C.~S.~Li, L.~L.~Yang and L.~G.~Jin,
  Phys.\ Rev.\  D {\bf 72} (2005) 074018;
  L.~L.~Yang, C.~S.~Li, Y.~Gao and J.~J.~Liu,
  Phys.\ Rev.\  D {\bf 73} (2006) 074017.


\bibitem{Zhang:2008yn}
  J.~J.~Zhang, C.~S.~Li, J.~Gao, H.~Zhang, Z.~Li, C.~P.~Yuan and T.~C.~Yuan,
  Phys.\ Rev.\ Lett.\  {\bf 102} (2009) 072001.


\bibitem{'tHooft:1972fi}
  G.~'t Hooft and M.~J.~G.~Veltman,
  Nucl.\ Phys.\  B {\bf 44} (1972) 189.


\bibitem{Catani:2002hc}
  S.~Catani, S.~Dittmaier, M.~H.~Seymour and Z.~Trocsanyi,
  Nucl.\ Phys.\  B {\bf 627} (2002) 189.


\bibitem{Collins:1978wz}
  J.~C.~Collins, F.~Wilczek and A.~Zee,
  Phys.\ Rev.\  D {\bf 18} (1978) 242.

\bibitem{Zhang:2009jj}
  J.~J.~Zhang, C.~S.~Li, J.~Gao and H.~X.~Zhu,
  in preparation.

\bibitem{Denner:1991kt}
  A.~Denner,
  Fortsch.\ Phys.\  {\bf 41} (1993) 307.


\bibitem{Ellis:2007qk}
  R.~K.~Ellis and G.~Zanderighi,
  JHEP {\bf 0802} (2008) 002.


\bibitem{Amsler:2008zzb}
  C.~Amsler {\it et al.}  [Particle Data Group],
  Phys.\ Lett.\  B {\bf 667} (2008) 1.

\bibitem{Pumplin:2002vw}
  J.~Pumplin, D.~R.~Stump, J.~Huston, H.~L.~Lai, P.~M.~Nadolsky and W.~K.~Tung,
  JHEP {\bf 0207} (2002) 012.

\bibitem{Frederix:2008hu}
  R.~Frederix, T.~Gehrmann and N.~Greiner,
  JHEP {\bf 0809} (2008) 122.

\end{thebibliography}
\end{document}